\documentclass[12pt]{report}
\usepackage{suthesis-2e}
\usepackage[top=3.5cm,bottom=3.5cm,left=3cm,right=3cm]{geometry}
\usepackage{epsfig}
\usepackage{amsthm}
\usepackage{amsfonts}
\usepackage{amssymb}
\usepackage{graphicx}
\usepackage{subfig}
\usepackage{esvect}
\usepackage{array}
\usepackage{commath}
\usepackage{chngcntr}
\usepackage{comment}
\usepackage{dcolumn}
\newcolumntype{d}[1]{D{.}{.}{#1}}    
\usepackage{siunitx}
\usepackage[T1]{fontenc}
\usepackage[english]{babel}
\usepackage{csquotes}
\usepackage{xcolor}
\newcommand{\Var}{\mathrm{Var}}

\usepackage[backend=biber]{biblatex} 
\usepackage{xurl}
\usepackage[hidelinks]{hyperref}

\usepackage[hyphenbreaks]{breakurl}

\begin{document}

\title{Deterministic and Random Perturbations of the Kepler Problem}
\author{Jesse Dimino}

\date{\today}
\dept{Mathematics}
\submitdate{May 2022}
\tablespagefalse
\principaladviser{Tobias Sch{\"a}fer}
\secondreader{Carlo Lancellotti}
\beforepreface
\prefacesection{Acknowledgements}
I would like to thank Professor Sch{\"a}fer for all of his help in making this paper possible through his advisement and guidance. I also wish to extend my appreciation to my family, friends, advisors, and mentors for their continued support. This work would not have been possible without them, and it is a product of their immeasurable reassurance and assistance.
\afterpreface

\begin{abstract}
    We investigate perturbations in the Kepler problem. We offer an overview of the dynamical system using Newtonian, Lagrangian and Hamiltonian Mechanics to build a foundation for analyzing perturbations. We consider the effects of a deterministic perturbation in the form of a first order relativistic correction which change bounded orbits from standard to precessing ellipses. We also consider the effects of stochastic perturbations with certain potentials and evaluate the analytical results of mean exit times using Monte Carlo simulations. 
\end{abstract}


\counterwithin{figure}{section}
\counterwithin{table}{section}
\counterwithin{equation}{section}
\counterwithin{theorem}{section}


\chapter{Introduction}
Johannes Kepler was a German mathematician and astronomer, who is most famous for his work on studying the orbits of the planets in the 17\textsuperscript{th} century. 
Through empirical observations, he outlined the following three natural laws to describe the orbits on planets in the Solar System.
\\
\\
1) Each planet moves in an elliptical orbit with the sun as a focus point.
\\
\\
2) The radial orbit of each planet sweeps out equal area in equal time.
\\
\\
3) The square period of revolution, $T$, is related to the cube of the major axis, $a$, by $T^2 = k a^3$ where $k$ is a constant for all planets.
\\
\\
Kepler was constrained by the development of mathematics available during his lifetime, and it was not until Sir Isaac Newton published his works on the Laws of Motion and the Universal Law of Gravitation, that these laws could be thoroughly understood and derived mathematically. Section \ref{sec:Newton} deals provides the derivation of the Kepler's laws and the equations of motion of the Kepler problem by utilizing Newtonian Mechanics. We then study the corresponding dynamics and compare the analytical solution to numerical simulations.

We also consider the effect of deterministic perturbations on the system. In Section \ref{sec:Lagrange}, we utilize Lagrangian mechanics to arrive at the analytical solution to the Kepler problem with a deterministic perturbation term that comes from the first order solution to the Einstein field equations in General Relativity. We then study the corresponding dynamics and fit our model to account for the precession of Mercury. 

We then transition to random perturbations. We utilize the action-angle coordinates to model our stochastic perturbations, so Section \ref{sec:Hamilton} gives a brief description of the system under Hamiltonian Mechanics to provide the basis for this coordinate system. We give an overview of random perturbations and stochastic differential equations in Section \ref{sec:SDE}. We then discuss the concept of exit times and our numerical schemes for simulating stochastic processes in Section \ref{sec:Exit}. Under certain conditions, the Kepler problem represents the semi classical limit of the nonlinear Schrödinger equation. In such cases, the propagation of solitons in an optical fiber are fundamentally similar to the orbits of the stochastically perturbed Kepler problem. Section \ref{sec:SKP} then deals with the derivation of the modulation equations for these stochastically pertubed potentials and the actual simulation of these processes. In particular, we consider a potential motivated from the equations of the harmonic oscillator to outline arrive at an analytical result for the mean exit time of the stochastic Kepler problem, and we verified this result using our discussed numerical methods.

\chapter{Classical Mechanics}
\section{Newtonian Formulation}
\label{sec:Newton}
Newton's Equation of Gravitation and Laws of Motion provide a sufficient basis for deriving Kepler's laws, and we derive the equations of motion of the Kepler problem utilizing Newtonian Mechanics. Similar derivations are outlined in \cite{MITNonLinDyn} and \cite{MITPlanMotion} which may be referenced for more detail.

To begin, consider Newton's Law of Gravitation,
\begin{equation}
    F_{G} = \frac{Gm_1m_2}{r^2}\hat{r}.
\end{equation}
Where $G$ is the gravitational constant, $m_1$ and $m_2$ are the masses of each body, $r$ is the magnitude of the distances between the centers of each body, and $\hat{r}$ is the unit vector pointing between the centers of each bodies defined as $\Vec{r}/r$.

To derive Kepler's laws, we may consider two masses, $m_1,m_2$, with coordinates $\vec{r}_1=(x_1,y_1,z_1)$, $\vec{r}_2=(x_2,y_2,z_2)$ respectively, experiencing mutual attraction in accordance with Newton's Law of Gravitation. The distance between $m_1, m_2$ is given by $\vec{r} = \vec{r_1} - \vec{r_2}$ and the center of mass of the system is $\vec{r}_{cm} = (m_1r_1 + m_2r_2)/(m_1+m_2)$. From Newton's Third Law it follows that
\begin{equation}
    \vec{F}_{1,2} = \frac{Gm_1m_2}{r^2}\hat{r} = -\vec{F}_{2,1}.
\end{equation}
Applying Newton's Second Law  to each equation leads to the following result
\begin{equation}
    \frac{\vec{F}_{1,2}}{m_1} - \frac{\vec{F}_{2,1}}{m_2} =\frac{d^2}{dt^2} \vec{r_1} - \frac{d^2}{dt^2} \vec{r_2} = \frac{d^2}{dt^2} \vec{r}.
\end{equation}
And by reapplying Newton's Third Law,  $\vec{F}_{1,2}$ factors out from the expression to obtain
\begin{equation}
    \vec{F}_{1,2}\left(\frac{1}{m_1}+\frac{1}{m_2}\right) = \frac{d^2}{dt^2} \vec{r}.
\end{equation}
We may then introduce the substitution for the reduced mass of the system
\begin{equation}
    \frac{1}{\mu} = \frac{1}{m_1} + \frac{1}{m_2} \implies \mu = \frac{m_1m_2}{m_1 + m_2} .
\end{equation}
Then by substitution, we obtain
\begin{equation}
    \vec{F}_{1,2} = \mu \frac{d^2}{dt^2} \vec{r}.
\end{equation}
This shows that the original two body problem can be rewritten as a one body problem with reduced mass $\mu$ and a position vector $\vec{r}$ with respect to a new central point. If we assume this system is acting only under the radially attractive force of gravity, $F$, the angular momentum, $L$, with respect to the origin is constant as
\begin{equation}
\begin{split}
    &\frac{dL}{dt} = \vec{r} \times \vec{F},\\
    &\frac{dL}{dt} = \lVert \vec{r} \rVert \lVert \vec{F} \rVert \sin{\pi} = 0.
\end{split}
\end{equation}

Thus angular momentum is conserved which implies that the motion takes place in a plane. As a result, a six-dimensional problem has been condensed into a two-dimensional problem. Standard polar coordinates $(r,\theta)$ will now be used to represent the reduced body problem. 
\\
\\
The other consequence of assuming that the system is isolated is the conservation of energy. It then follows that the total energy of the system is simply the sum of the kinetic and potential energy
\begin{equation}
    E = E_K + E_P = \frac{1}{2}\mu v^2 - \frac{Gm_1m_2}{r}.
\end{equation}
 We have that $\vec{v} = v_{rad}\hat{r} + v_{tan} \hat{\theta}$ and by definition $v_{rad} = \dot r$ and $v_{tan} = r\dot\theta$, and we obtain
\begin{equation}
    v^2 = v_{rad}^2+v_{tan}^2 = \left(\frac{dr}{dt}\right)^2 + \left(r\frac{d\theta}{dt}\right)^2.
\end{equation}
Then a simple substitution into the equation for total energy gives
\begin{equation}
    E = \frac{1}{2}\mu\left(\left(\frac{dr}{dt}\right)^2 + \left(r\frac{d\theta}{dt}\right)^2\right)-\frac{Gm_1m_2}{r}.
\end{equation} 
It also follows that an equivalent expression for the angular momentum  with respect to the center of mass is
\begin{equation}
    L = r\mu v_{tan} \sin{\frac{\pi}{2}} = r\mu r\frac{d\theta}{dt} = r^2 \mu \frac{d\theta}{dt}. 
\end{equation}
Applying Newton's Second Law to the reduced system gives the following result
\begin{equation}
        \frac{-Gm_1m_2}{r^2}\hat{r} = \mu\left(\frac{d^2r}{dt^2}-r\left(\frac{d\theta}{dt}\right)^2\right)\hat{r}.
\end{equation}
There is a direct implication that the orbit is no longer strictly circular due to the radial acceleration term along with the standard centripetal acceleration term. It is then possible to note the following,
\begin{equation}
    L^2 = \mu^2r^4\left(\frac{d\theta}{dt}\right)^2 \implies \left(\frac{d\theta}{dt}\right)^2 = \frac{L^2}{\mu^2r^4}.
\end{equation}
Via a simple substitution 
\begin{equation}
\label{eq:KeplerDiffEq}
    \frac{d^2r}{dt^2} = \frac{L^2}{\mu^2 r^3} - \frac{Gm_1m_2}{\mu r^2}.
\end{equation}
We may then introduce the change of variables, $u = 1/r$, to obtain 
\begin{equation}
    \frac{dr}{dt} = \frac{dr}{du}\frac{du}{d\theta}\frac{d\theta}{dt} = -\frac{1}{u^2}\frac{du}{d\theta}\frac{Lu^2}{\mu}= -\frac{du}{d\theta}\frac{L}{u}.
\end{equation}
This allows us to take the second derivative with respect to $t$ to obtain
\begin{equation}
\frac{d^2r}{dt^2} = \frac{d}{dt}\frac{dr}{dt} = \frac{d}{d\theta}\frac{dr}{dt}\frac{d\theta}{dt} = \frac{d}{d\theta}\frac{-du}{d\theta}\frac{L}{\mu}\frac{Lu^2}{\mu} = -\frac{d^2u}{d\theta^2}\frac{L^2u^2}{\mu^2}.
\end{equation}
Substituting back into equation \ref{eq:KeplerDiffEq} then yields
\begin{equation}
\label{eq:KepU}
\frac{d^2u}{d\theta^2} + u = \frac{\mu Gm_1m_2}{L^2}.
\end{equation}
This equation is equivalent to the inhomogeneous harmonic oscillator, with a general solution of the form
\begin{equation}
    u = u_0 + A\cos{(\theta-\phi)}.
\end{equation}
Where $A$ and $\phi$ are constants determined by the form of the orbit. We may then define 
\begin{equation}
\begin{split}
    &u_0 = \frac{\mu Gm_1m_2}{L^2} \implies\\
    &r_0 = \frac{1}{u_0} = \frac{L^2}{\mu Gm_1m_2}.
\end{split}
\end{equation}
We then obtain the following solution
\begin{equation}
r = \frac{r_0}{1+r_0A\cos{(\theta-\phi)}}.
\end{equation}
There are now two degrees of freedom to choose the constants $A,\phi$. At this point, it is worth introducing the eccentricity, $\epsilon$ , of the orbit \cite{MITPlanMotion} which is defined as
\begin{equation}
    \epsilon = \sqrt{1+\frac{2EL^2}{\mu G^2 m_1^2 m_2^2}}.
\end{equation}
Set $A=\epsilon/r_0$ and $\phi = \pi$ then $\cos{(\theta-\phi)} = \cos{(\theta-\pi)} = -\cos{(\theta)}$, to obtain
\begin{equation}
    r = \frac{r_0}{1-\epsilon\cos{\theta}}.
\end{equation}
From \cite{MITPlanMotion}, we can note the following,
\begin{equation}
    \begin{split}
        &L = \sqrt{\mu G m_1 m_2 r_0},\\
        &E = \frac{Gm_1m_2 (\epsilon^2-1)}{2r_0},\\
        &r = r_0 + r\epsilon \cos{\theta}.
    \end{split}
\end{equation}
We can see that $r$ has the general form of a conic section in polar coordinates. With all of that being noted there is now a sufficient basis to derive Kepler's laws analytically.

Recall that Kepler's First law states that the planets move in an elliptical orbit with the sun as a focus point. An ellipse is a particular case of a conic section with the restriction that $0 \leq \epsilon < 1$. For two bodies in a bound orbit the the potential energy due to the mutual force of attraction must be stronger than the kinetic energy, otherwise one of the bodies would be able to escape orbit. Therefore the total energy of the system must be negative. The factor $(Gm_1m_2)/(2r_0)$ is positive, thus the sign of the energy term is only dependent upon $\epsilon$. It's then trivial to note that energy is only negative when $0 \leq \epsilon < 1$. Thus it follows that two bodies in a bounded orbit do so in an ellipse. To show that the sun is located at one of the focus points of the ellipse, recall that the center of mass of the system is given by
\begin{equation}
\begin{split}
    &\vec{r_{cm}} = \frac{m_1\vec{r_1} + m_2\vec{r_2}}{m_1+m_2},\\
    &\vec{r} = \vec{r_1} -\vec{r_2}.
\end{split}
\end{equation}
It then follows
\begin{equation}
    \vec{r_1}' = \vec{r_1}-\vec{r_{cm}} = \vec{r_1}-\frac{m_1\vec{r_1} + m_2\vec{r_2}}{m_1+m_2} = \frac{m_2(\vec{r_1}-\vec{r_2})}{m_1+m_2} = \frac{\mu}{m_1}\vec{r} \ , \ \vec{r_2}' = -\frac{\mu}{m_2}\vec{r}.
\end{equation}
Thus each body undergoes motion around the center of mass in the same vein that the reduced body does around the central point, just adjusted by a factor of $\mu/m_i$. Consider the case where $m_2 \gg m_1$, such as with the sun's mass relative to the mass of all the planets, in this case $\mu$ is approximately the smaller mass 
\begin{equation}
    \mu = \frac{m_1m_2}{m_1+m_2} \approx m_1.
\end{equation}
It then follows that
\begin{equation}
    \vec{r_1}' = \frac{\mu}{m_1}\vec{r} \approx \frac{m_1}{m_1}\vec{r} = \vec{r}.
\end{equation}
Similarly,
\begin{equation}
    \vec{r_2}' = -\frac{\mu}{m_2}\vec{r} \approx - \frac{m1}{m_2}\vec{r} \approx 0.
\end{equation}
Thus $m_2$ is approximately stationary with $m_1$ orbiting around it when $m_2 \gg m_1$. As this was motivated by the example of the sun's relative mass to the planets, each planet orbits the sun in an ellipse with the sun at one of the focus points thus proving Kepler's First law. 
\\
\\
Now recall that Kepler's second law states that the radial orbit of each planet sweeps out equal area in equal time. As shown previously, the motion takes place in a plane thus the area swept out by the radial vector is half the area of the parallelogram formed by $\vec{r}$ and $\vec{dr}$.
\begin{equation}
\label{eq:KeplerLaw2}
    dA = \frac{1}{2}\lvert \vec{r} \times \vec{dr} \rvert = \frac{1}{2}\lvert \vec{r} \times \frac{\vec{dr}}{dt}dt \rvert = \frac{L}{2\mu}dt.
\end{equation}
We can note that $L/(2\mu)$ is constant, thus the radial vector sweeps out equal areas in equal time, proving Kepler's Second Law.
\\
\\
Recall that Kepler's third law is the statement, 
\begin{equation}
    T^2 = ka^3.
\end{equation}
where $k$ is a constant for all planets, $T$ is the orbital period, and $a$ is the semimajor axis of the ellipse. Equation \ref{eq:KeplerLaw2} can be rewritten in the following form,
\begin{equation}
    \frac{dA}{dt} = \frac{L}{2\mu}.
\end{equation}
This can then be integrated to obtain
\begin{equation}
    \frac{2\mu}{L} \oint dA = \int_0^Tdt.
\end{equation}
Where $\oint dA$ is simply the area of the ellipse which has the standard formula,
\begin{equation}
    A = \pi a b,
\end{equation}
where $a$ and $b$ are the semimajor and semiminor axes of the ellipse respectively. Thus equation \ref{eq:KeplerLaw2} reduces to:
\begin{equation}
   T = \frac{2\mu \pi a b}{L}.
\end{equation}
Squaring both sides gives
\begin{equation}
    T^2 = \frac{4\pi^2 \mu^2 a^2 b^2}{L^2}.
\end{equation}
The angular momentum can then be rewritten in terms of the semimajor axis using $L = \sqrt{\mu G m_1 m_2 a(1-\epsilon^2)}$, and another substitution can be made to rewrite the semiminor axis using $b = \sqrt{(aL^2)/(\mu Gm_1 m_2)}$, as outlined in \cite{MITPlanMotion}, to obtain the following result
\begin{equation}
    T^2 = \frac{4 \pi^2 a^3}{G(m_1+m_2)}.
\end{equation}
In the case that $m_2 \gg m_1$ such as with the sun and the planets this approximately becomes
\begin{equation}
    T^2 = \frac{4\pi^2 a^3}{Gm_2}.
\end{equation}
We then define $k = (4\pi^2)/(Gm_2)$ and substitute to obtain
\begin{equation}
    T^2 = ka^3.
\end{equation}
We have that $k$ is constant for all planets, proving Kepler's third law.


\section{Dynamics of the Kepler Problem}
\label{sec:KepDyn}

The Kepler Problem is governed by a second order ordinary differential equation, and it's worth noting that an explicit solution in the form $r(t)$ cannot be obtained without special functions, but we can find a solution for $r(\theta)$. The system in general is 4-dimensional, but if we consider $\theta$ as an independent variable instead of a function of time, we can reduce it to 2 dimensions and more comfortably analyze the underlying dynamics of the system. We can also consider $u(\theta)$ as the equations are much simpler than the corresponding representation in $r(\theta)$. We can begin by rewriting equation \ref{eq:KepU} into a system of first order differential equations as follows

\begin{equation}
\begin{split}
        &u' = v,\\
        &v' = -u + u_0.
\end{split}
\end{equation}
And we can see that there is a single fixed point at $(u_0,0)$ which corresponds to being at the center of the orbit with no initial velocity. We can then classify this fixed point by utilizing linear stability analysis. We may note that the corresponding Jacobian of the system is 

\begin{equation}
    \mathcal{J}(u,v) = 
    \begin{pmatrix}
    0 & 1\\
    -1 & 0
    \end{pmatrix}.
\end{equation}

This has eigenvalues of $\pm i$ for any choice of inputs. These eigenvalues correspond to a center which is to be expected, and since we assumed that the system is conservative, we can be assured that the linearization around that fixed point holds, and this analysis represents the dynamics of bounded orbits that are sufficiently far from the separatrix in the original equations.

\begin{figure}[ht]%
    \centering
    \subfloat[\centering Analytical Solution]{{\includegraphics[height = 5cm, trim={0 0 0 0 }, clip]{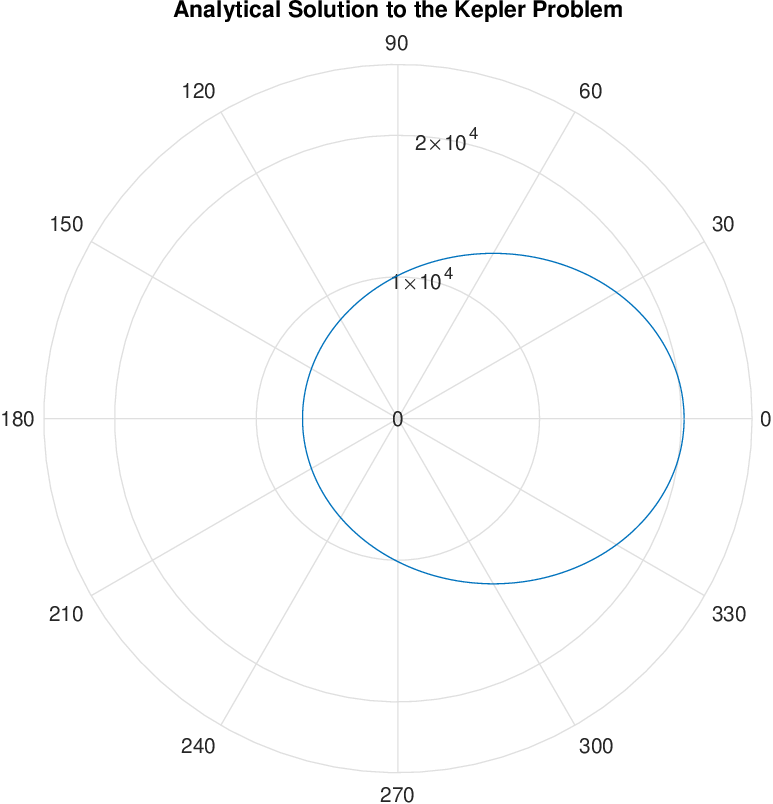}}}%
    \qquad
    \subfloat[\centering Numerical Solution]{{\includegraphics[height = 5cm, trim={0 0 0 0 }, clip]{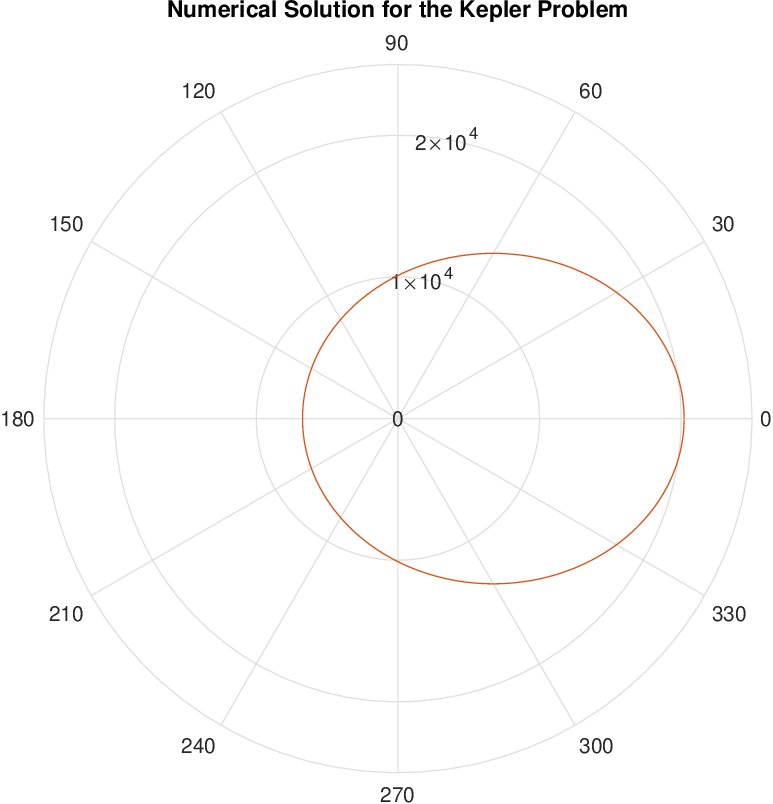}}}%
    \caption{Analytical vs numerical solution to the Kepler Problem.}%
    \label{fig:KeplerSolutions}%
\end{figure}

The general solution to the Kepler problem is given by
\begin{equation}
    r = \frac{r_0}{1-\epsilon\cos{\theta}}.
\end{equation}
For computational convenience, we can consider reasonable values to plot the resulting orbit. In particular we use $G = 1, m_1 = 0.01, m_2 = 1, L = 1, \epsilon = 0.5$ and we have that $\mu = (m_1m_2)/(m_1 + m_2), r_0 = L^2/(\mu Gm_1m_2)$. We will consider the analytical solution, and the numerical approximation using the classical Runge-Kutta scheme (RK4) using a step size $h = 0.01$ and the results are shown in Figure \ref{fig:KeplerSolutions}. The classical Runge-Kutta method is a 4th order numerical method for solving ordinary differential equations which works by initializing the scheme with the given conditions and then utilizing a weighted average of the derivative's values at discrete intervals to approximate a new value of the solution. The scheme may then be iterated using this newly approximated value to trace out the solution up to some specified final point. A more thorough explanation of the RK4 scheme and its convergence is outlined in \cite{Numerics}.

\begin{figure}[ht]
    \centering
    \includegraphics[height = 5cm, trim={0 0 0 0 }, clip]{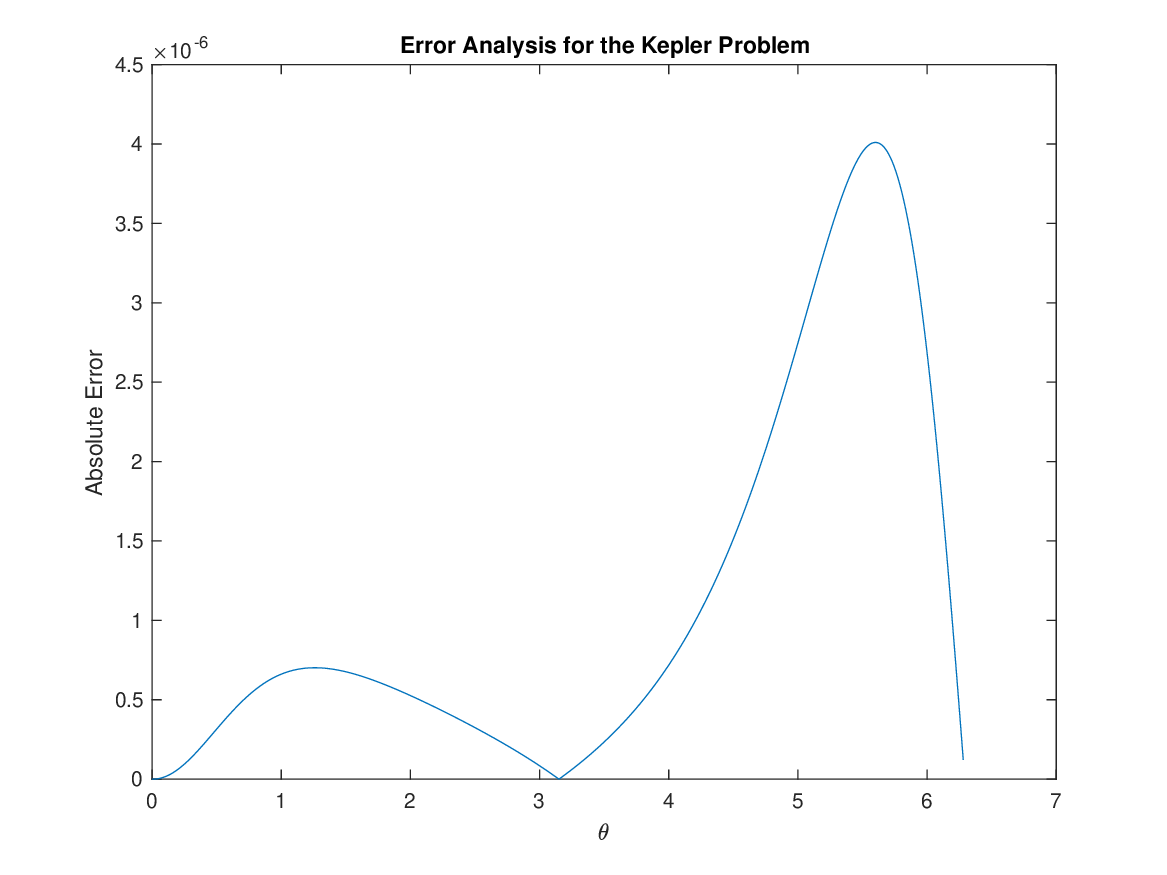}
    \caption{Error Analysis for the Kepler Problem.}
    \label{fig:KeplerError}
\end{figure}

The RK4 scheme has an error $\propto h^4$, so if the scheme is working as intended we expect to see an error roughly proportional $10^{-8}$ times the scale of the units that we are using. the graph of the error between the numerical and analytical solutions is given in figure \ref{fig:KeplerError}.

The maximum order is on the scale of $10^{-6}$ which is an indication that the RK4 scheme works well to approximate this solution, but we still have to ensure that it is truly capturing the dynamics of the problem, and to that end we may consider plotting the phase space using both the analytical and numerical solutions. 

\begin{figure}[ht]%
    \centering
    \subfloat[\centering Analytical Solution]{{\includegraphics[height = 5cm, trim={0 0 0 0 }, clip]{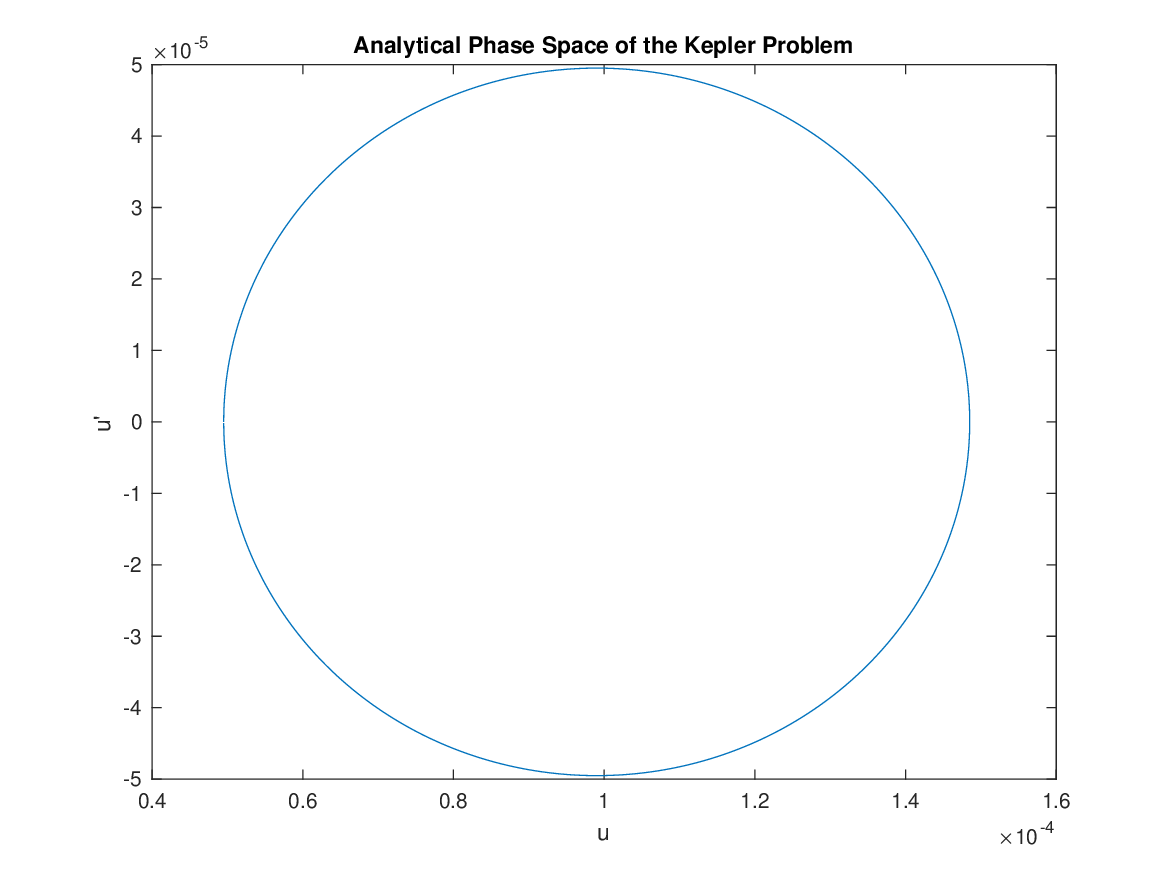}}}%
    \qquad
    \subfloat[\centering Numerical Solution]{{\includegraphics[height = 5cm, trim={0 0 0 0 }, clip]{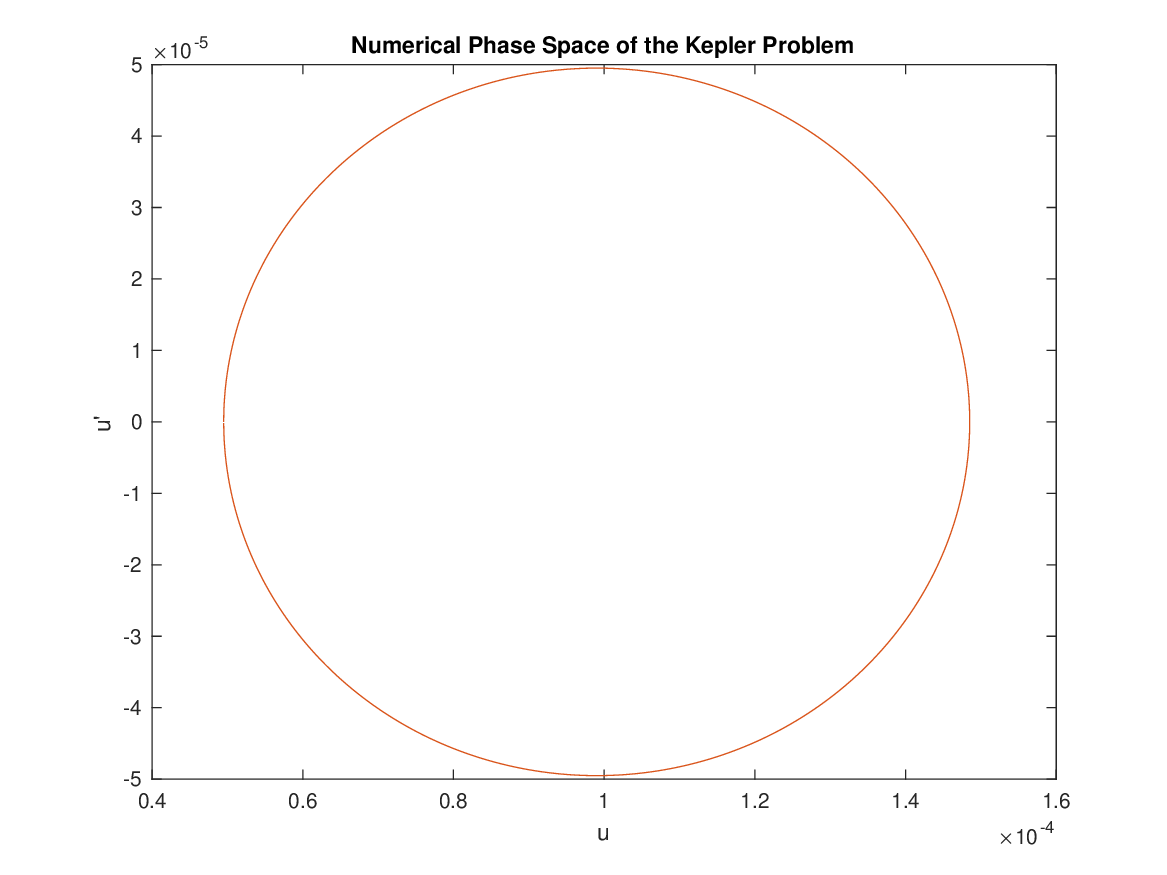}}}%
    \caption{Phase space of the Kepler Problem.}%
    \label{fig:KeplerPhaseSpace}%
\end{figure}
The plots are extremely similar which gives confidence that the RK4 scheme is capturing the dynamics of the system. The order is on $10^{-5}$ which is a consequence of the simplified values that we used. When applying the analysis to any realistic example, the expected values are many orders of magnitude larger, but to ensure our computations were accurate and avoid machine error, we opted to use simpler values. In principle we also could have non-dimensionalized the equations to make them more suitable for numerics, but the coefficients still have physical significance and we opted to keep them in order to ensure this information was preserved.


\section{Lagrangian Mechanics and Deterministic Perturbations}
\label{sec:Lagrange}
The standard Kepler Potential is
\begin{equation}
    V(r) = \frac{-Gm_1m_2}{r}.
\end{equation}
To build on the dynamics of the Kepler Problem, we can consider a perturbed potential as follows:

\begin{equation}
     V(r) = \frac{-Gm_1m_2}{r}\left( 1 + \alpha \frac{Gm_2}{rc^2}\right).
\end{equation}
This added term to the standard Kepler potential represents the first order correction from the Einstein field equations which form the basis of General Relativity. The term $\alpha$ is a dimensionless parameter and $c$ is the speed of light. To arrive at the equations of motion, we may consider the Lagrangian of the system. Lagrangian mechanics seeks to describe the system using generalized velocities instead of describing motion under a system of forces. The main idea is that the action functional of the system derived from the Lagrangian must remain stationary, so we can then calculate the corresponding equations of motion from the Euler-Lagrange equation. For a more comprehensive overview of Lagrangian Mechanics see \cite{Mechanics}. The Lagrangian of a system is typically defined as follows,
\begin{equation}
    \mathcal{L} = T - V.
\end{equation}
This is simply the difference of the kinetic and potential energies. In the perturbed Kepler case this becomes
\begin{equation}
    \mathcal{L} = T - V =\frac{1}{2}\mu[(\frac{dr}{dt})^2 + r^2(\frac{d\theta}{dt})^2]+\frac{Gm_1m_2}{r}\left( 1 + \alpha \frac{Gm_2}{rc^2}\right).
\end{equation}
To arrive at the desired equation of motion, we may utilize the Euler-Lagrange equation,
\begin{equation}
    \frac{\partial \mathcal{L}}{\partial r} - \frac{d}{dt}\frac{\partial \mathcal{L}}{\partial \dot r} = 0.
\end{equation}
The solution to this equation will be a functional that minimizes the action of the system, and in classical mechanics that corresponds to the canonical path. With the relativistic perturbation, we obtain the equation
\begin{equation}
    \mu \ddot r = \mu r \dot \theta^2 -\frac{Gm_1m_2}{r^2} - 2\alpha \frac{G^2m_1m_2^2}{r^3c^2}.
\end{equation}
If we once again utilize the change of variables $u = 1/r$, we can rewrite the equations as follows
\begin{equation}
    \frac{d^2u}{d\theta^2} + \left( 1 - \frac{2 \alpha G^2 m_1m_2^2\mu}{L^2c^2}\right)u = \frac{Gm_1m_2\mu}{L^2}.
\end{equation}
And we can introduce the substitutions
\begin{equation}
    \begin{split}
        &\xi = \left( 1 - \frac{2 \alpha G^2 m_1m_2^2\mu}{L^2c^2}\right),\\
        &q = \frac{Gm_1m_2\mu}{L^2}.
    \end{split}
\end{equation}
We may then obtain
\begin{equation}
\label{eq:KepUPerturbed}
    \frac{d^2u}{d\theta^2} + \xi u = q.
\end{equation}
This is the parameterized inhomogeneous harmonic oscillator. This system has a solution of the form
\begin{equation}
    u = A\cos{(\sqrt{\xi}\theta - \phi)} + \frac{q}{\xi}.
\end{equation}
We may now set $u_0 = q/\xi$, $\phi = \pi$, $A = \epsilon/r_0$, and we may note that $r_0 = 1/u_0 = \xi/q$ to rewrite the solution as follows
\begin{equation}
    u = \frac{1}{r_0}\left(1- \epsilon\cos{\left(\sqrt{\xi}\theta\right)} \right).
\end{equation}
And we may finally rewrite the solution in terms of $r$, to obtain
\begin{equation}
    r = \frac{r_0}{1-\epsilon \cos({\sqrt{\xi}\theta})}.
\end{equation}
This represents a modified solution to the equation found with Newtonian Mechanics. If $0 \leq \epsilon < 1$ and $\sqrt{\xi} \neq 1 $ then the orbits represent a precessing ellipse.


\section{Dynamics of the Perturbed Kepler Problem}
The perturbed Kepler problem is governed by a second order differential equation, and we may analyze the dynamics in a similar manner as in Section \ref{sec:KepDyn}. We may first consider representing equation \ref{eq:KepUPerturbed} as a system of first order differential equations.
\begin{equation}
    \begin{split}
        &u' = v,\\
        &v' = -\xi  u + q.
    \end{split}
\end{equation}
We can see that there is a single fixed point at $(q/\xi,0)$. Since we previously defined $u_0 = q/\xi$, we can see that the fixed point is at $(u_0,0)$ which corresponds to the unperturbed case. We can note that the corresponding Jacobian of the system is
\begin{equation}
    \mathcal{J}(u,v) = 
    \begin{pmatrix}
    0 & 1\\
    -\xi & 0
    \end{pmatrix}.
\end{equation}
This has eigenvalues of $\pm \xi i$, and this corresponds to a center although it has been scaled by a factor of $\xi$ from the standard Kepler potential. The system is conservative which ensures that the linearization around the fixed point holds, so higher order effects will not disrupt the stability of the orbits.

\begin{figure}[ht]%
    \centering
    \subfloat[\centering Analytical Solution]{{\includegraphics[height = 5cm, trim={0 0 0 0 }, clip]{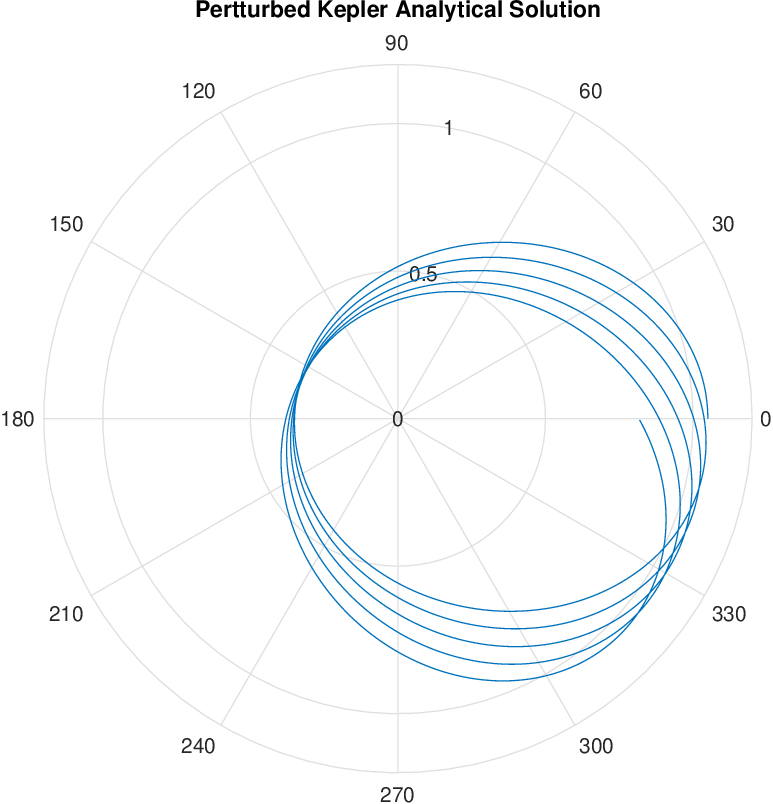}}}%
    \qquad
    \subfloat[\centering Numerical Solution]{{\includegraphics[height = 5cm, trim={0 0 0 0 }, clip]{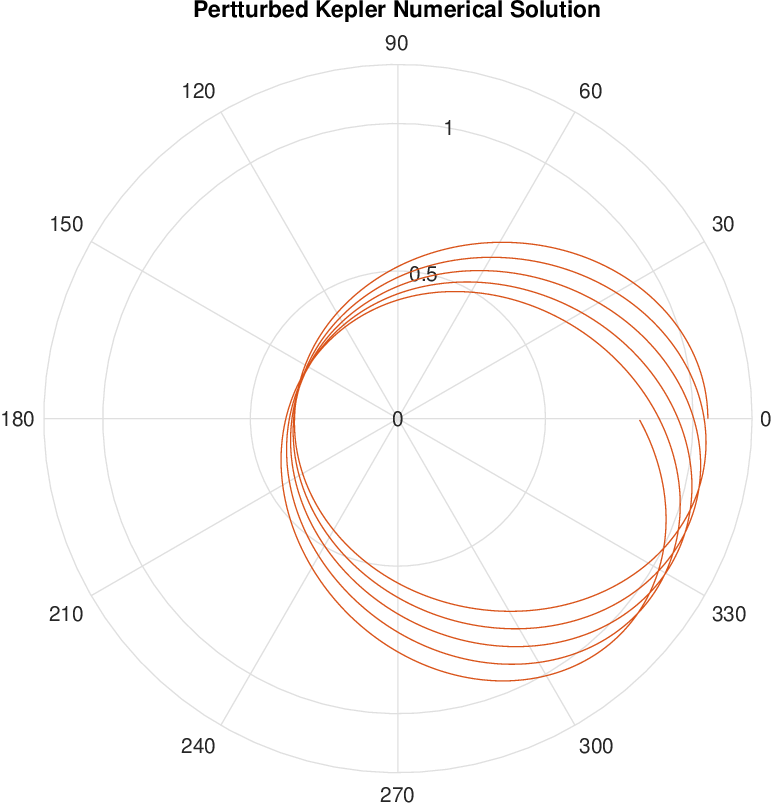}}}%
    \caption{Analytical vs numerical solution to the Perturbed Kepler Problem.}%
    \label{fig:PerturbedKeplerSolutions}%
\end{figure}
To ensure that the numerics are accurate, we may also compare the absolute error between the two solutions.
\begin{figure}[ht]
    \centering
    \includegraphics[height = 5cm, trim={0 0 0 0 }, clip]{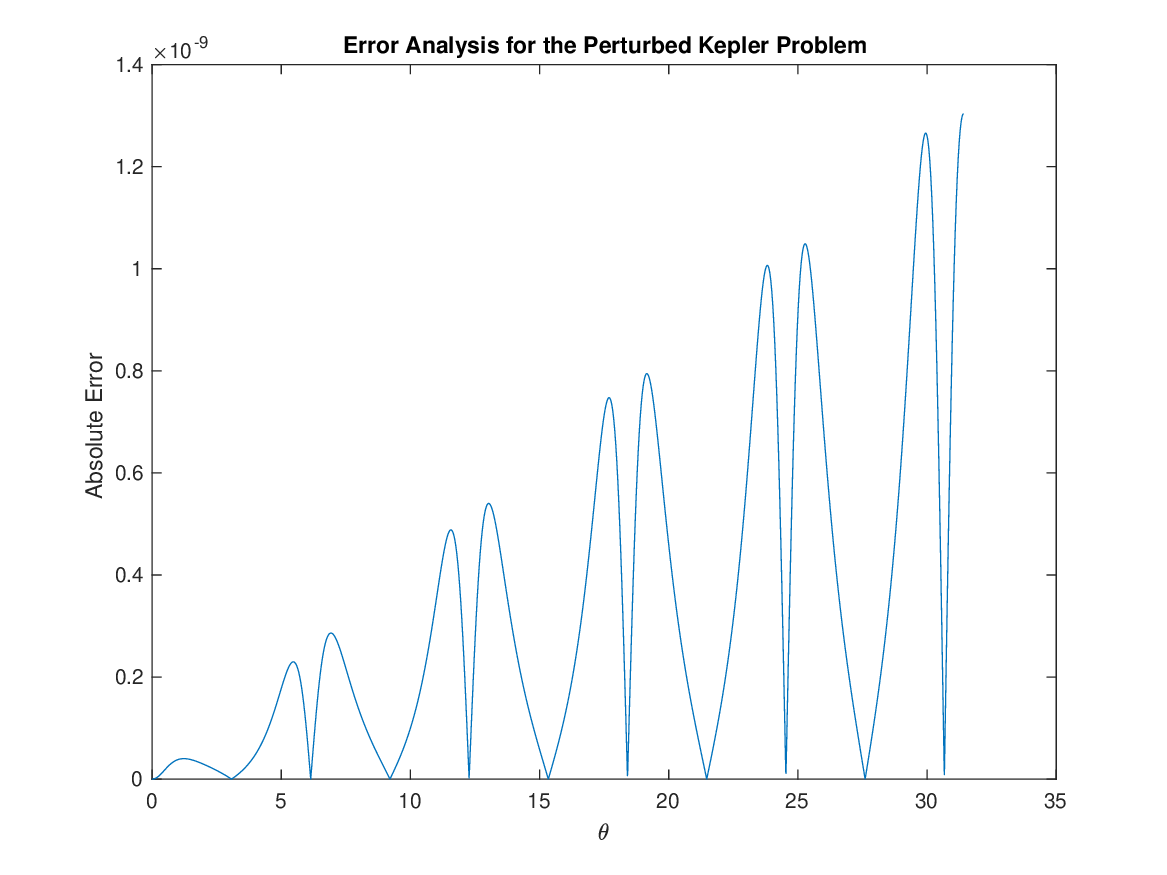}
    \caption{Error Analysis for the Perturbed Kepler Problem.}
    \label{fig:PerturbedKeplerError}
\end{figure}
The error is on a small enough order to provide a sufficient approximation for small timescales. We may then check the phase space to make sure that the scheme is capturing the dynamics as well.
\begin{figure}[ht]%
    \centering
    \subfloat[\centering Analytical Solution]{{\includegraphics[height = 5cm, trim={0 0 0 0 }, clip]{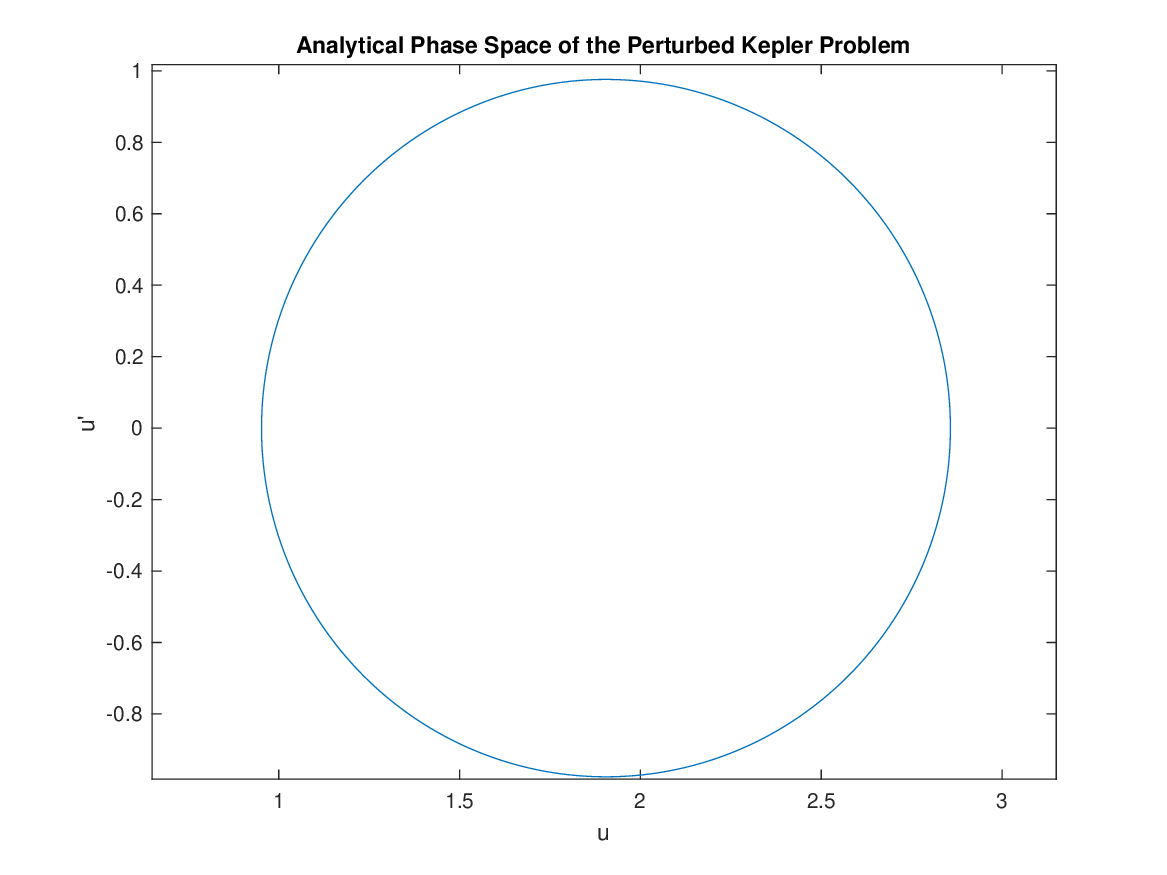}}}%
    \qquad
    \subfloat[\centering Numerical Solution]{{\includegraphics[height = 5cm, trim={0 0 0 0 }, clip]{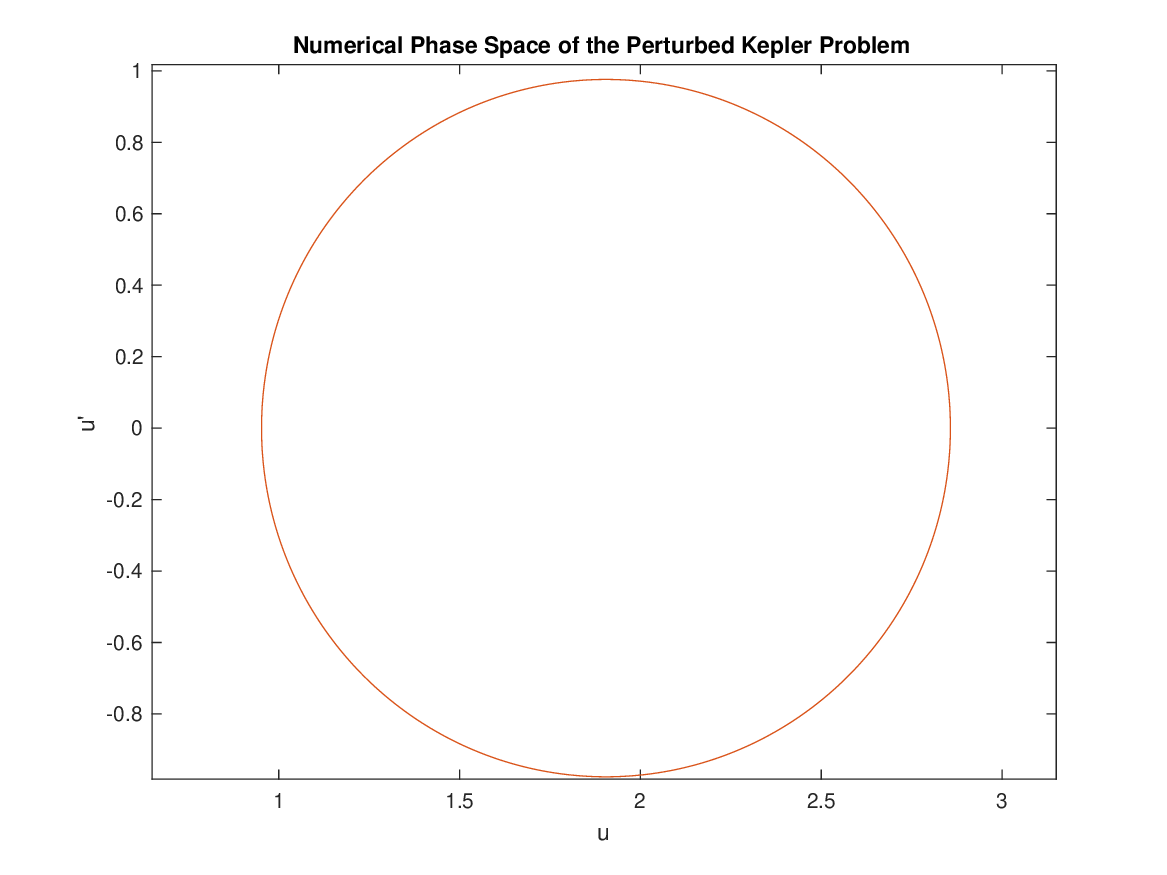}}}%
    \caption{Phase space of the Perturbed Kepler Problem.}%
    \label{fig:PerturbedKeplerPhaseSpace}%
\end{figure}
And the two phase space plots show that the numerics are accurately capturing the dynamics of the perturbed Kepler problem.


\section{The Precession of Mercury}
One of the classical tests of General Relativity was calculating the anomalous precession of Mercury's orbit. It was observed that Mercury's perihelion advances at a rate of 574 seconds of arc per century, but Newtonian mechanics and classical perturbation theory predicted a precession of 531 seconds of arc per century \cite{aether}. This anomalous 43" per century was correctly predicted by General Relativity without any modifications, and it is considered one of the theory's major successes. 

In most cases, the higher order solutions to the Einstein field equations are used for predictions, but they have the disadvantage that no analytical solution exists when those corrections are added. If we simply use the first order correction, it is still possible to formulate an analytical solution, and we can fit our parameters to account for observable relativistic effects with a tolerable margin of error. 

Recall first order relativistic Kepler potential has the form
\begin{equation}
    V(r) = \frac{-Gm_1m_2}{r}\left( 1 + \alpha \frac{Gm_2}{rc^2}\right).
\end{equation}
We defined the following variables,
\begin{equation}
    \begin{split}
        &\xi = \left( 1 - \frac{2 \alpha G^2 m_1m_2^2\mu}{L^2c^2}\right),\\
        &q = \frac{Gm_1m_2\mu}{L^2},\\
        &r_0 = \frac{\xi}{q}.
    \end{split}
\end{equation}
And the general solution is given by
\begin{equation}
     r = \frac{r_0}{1-\epsilon \cos({\sqrt{\xi}\theta})}.
\end{equation}
We may fit $\alpha$ to account for the anomalous precession, and we note that the constants of Mercury's orbit outlined in \cite{planets}. By definition the first perihelion occurs at $\theta = 0$, and the next perihelion occurs at
\begin{equation}
\label{eq:perihelion}
    \sqrt{\xi} \theta = 2\pi \implies \theta =\frac{2\pi}{\sqrt{\xi}}.
\end{equation}
We may introduce a change of variables such that $\beta = (\alpha G^2 m_1m_2^2\mu)/(L^2c^2) $ and $\xi = 1-2\beta$. We may the following identity as $\beta$ is strictly real valued \cite{Theory},
\begin{equation}
    \eta =\frac{2\pi}{\sqrt{1-2\zeta}} = 2\pi + 2\pi\zeta.
\end{equation}
By applying this result to equation \ref{eq:perihelion}, we obtain
\begin{equation}
    \theta = 2\pi + 2\pi\beta.
\end{equation}
Now in a standard ellipse, the second perihelion occurs at $2\pi$, so the anomalous precession, $\delta$ , is simply the difference from $2\pi$, therefore
\begin{equation}
    \delta = 2\pi \beta \implies \frac{\delta}{2\pi} = \beta \implies \frac{\delta}{2\pi} = \frac{\alpha G^2 m_1m_2^2\mu}{L^2c^2} \implies \alpha = \frac{\delta L^2 c^2}{2\pi G^2 m_1 m_2^2 \mu}.
\end{equation}
The value of $\delta$ for Mercury is known to be 43 seconds of arc per century or equivalently $5.0162*10^{-7}$ radians per revolution, and using this value to calculate $\alpha$ leads to the result
\begin{equation}
    \alpha = 3.1330.
\end{equation}
To verify the value of $\alpha$, it is also worth looking at the anomalous precession of Venus which is known to be 8.6 seconds of arc per century, or equivalently $2.5723*10^{-7}$ radians per revolution. By using the values which correspond to the physical constants for Venus in \cite{planets}, it is possible to calculate an expected anomalous precession of $2.6862*10^{-7}$ radians per revolution which agrees to within $4.5\%$ of the measured value. We can see that this value of $\alpha$ is suitable for modeling the motion of planets in the solar system. Therefore the first order correction provides a fairly accurate scheme to calculate relativistic precession while still maintaining an explicit solution; thus the reduced correction provides a good intermediary between Newtonian mechanics and the higher order correction more commonly used in General Relativity.


\section{Hamiltonian Mechanics}
\label{sec:Hamilton}
Another way to derive the equations of motion of the Kepler Problem is through Hamiltonian mechanics. Hamiltonian mechanics developed in the 1830s as a reformulation of Lagrangian mechanics. The Lagrangian of the system is generally defined as the difference between the kinetic and the potential energies whereas the Hamiltonian is defined by their sum, which is equivalent to the total energy in conservative systems. Hamiltonian mechanics differs from Lagrangian mechanics by considering generalized momenta in the system instead of velocities, but it can be used to explain the same physical systems. A more thorough introduction to Hamiltonian mechanics is covered in \cite{Mechanics}. The Hamiltonian of the Kepler problem is denoted by 
\begin{equation}
    \mathcal{H} = T + V =\frac{1}{2}\mu[(\frac{dr}{dt})^2 + r^2(\frac{d\theta}{dt})^2]-\frac{Gm_1m_2}{r}.
\end{equation}
The generalized momenta in Hamiltonian mechanics are defined by the following equation,
\begin{equation}
    P_i = \frac{\partial \mathcal{L}}{\partial \dot q_i}.
\end{equation}
There are two degrees of freedom in the Kepler problem, $r,\theta,$ so we consider the linear momentum and the radial momentum which can be expressed as
\begin{equation}
    \begin{split}
        &P_r = \frac{\partial \mathcal{L}}{\partial \dot r} = \mu \dot r,\\
        &P_{\theta} = \frac{\partial \mathcal{L}}{\partial \dot \theta} = \mu r^2 \dot \theta.
    \end{split}
\end{equation}
These are the classical linear and angular momenta that were derived with Newtonian Mechanics in Section \ref{sec:Newton}. Noting that $T = KE_r + KE_{\theta}$ it is then possible to write $T = P_r^2/(2\mu) + P_{\theta}^2/(2\mu r^2)$ and substitute into the Hamiltonian to obtain the following,
\begin{equation}
    \mathcal{H} = \frac{P_r^2}{2\mu} + \frac{P_{\theta}^2}{2\mu r^2} - \frac{Gm_1m_2}{r}.
\end{equation}
Then to derive the corresponding equations of motion, it is necessary to look at Hamilton's equations,
\begin{equation}
    \begin{split}
        &\frac{dq_i}{dt} = \frac{\partial H}{\partial P_i},\\
        &\frac{dP_i}{dt} = -\frac{\partial H}{\partial q_i}.
    \end{split}
\end{equation}
It is worth noting that the solution to the Lagrangian was an $n$\textsuperscript{th} order differential equation whereas the solution to Hamilton's Equations are $2n$ first order differential equations
\begin{equation}
    \begin{split}
        &\frac{dr}{dt} = \frac{\partial H}{\partial P_r} = \frac{P_r}{\mu},\\
        &\frac{dP_r}{dt} = -\frac{\partial H}{\partial r} = \frac{P_\theta^2}{\mu r^3}-\frac{Gm_1m_2}{r^2},\\
        &\frac{d\theta}{dt} = \frac{\partial H}{\partial P_{\theta}} = \frac{P_\theta}{\mu r^2},\\
        &\frac{dP_{\theta}}{dt} = -\frac{\partial H}{\partial \theta} = 0.
    \end{split}
\end{equation}
The equations are consistent with what was derived using Newtonian and Lagrangian mechanics which show that the formulations are identical for the Kepler problem.


\section{Action Angle Coordinates}
We can choose a specific coordinate system for the Kepler problem to simplify the representation of the action of the system. Our goal is to make the motion in phase space as simple as possible while leaving its dynamics untouched. This can be done through the use of canonical transformations. Cyclic variables are variables that do not explicitly appear in the Hamiltonian, and they are the easiest to use as a result. For some generalized cyclic coordinate, $q_i$, the following holds for its conjugate momentum: $\dot p_i = 0 \implies p_i = \alpha_i$. If the Hamiltonian is strictly a function of these cyclic variables and it is conserved (i.e. $\frac{\partial \mathcal{H}}{\partial t} = \frac{d \mathcal{H}}{dt}=0$), then $\mathcal{H}(\alpha_1,...,\alpha_n)$ is constant in time. The resulting equations for the general paths are linear in time as
\begin{equation}
    \dot q_i = \frac{\partial H}{\partial p_i} = \omega_i(\alpha_i) \implies q_i = \omega_it + \delta_i.
\end{equation}
We wish to find a set of canonical transformations, $(q_i,p_i)\rightarrow (Q_i,P_i)$, to make as many variables cyclic as possible. Ideally, we may construct an equivalence such that $H(q,p,t) = K(Q,P)$. To find these canonical transformations, we need a suitable generating function, $F$, such that
\begin{equation}
    p_i\dot q_i - H = P_i \dot Q_i - K + \dot F. 
\end{equation}
There are several cases to consider, but the one of most interest to this problem is the $F_2$ type generating function \cite{Mechanics}
\begin{equation}
    F = F_2(q,P,t) - P_iQ_i.
\end{equation}
It then follows that
\begin{equation}
\begin{split}
     &p_i\dot q_i - H - P_i \dot Q_i + K = \dot F_2 - \dot P_iQ_i - P_i \dot Q_i  \implies \\
    & p_i\dot q_i - H  + K + \dot P_iQ_i = \frac{\partial F_2}{\partial t} + \frac{\partial F_2}{\partial q_i}\dot q_i + \frac{\partial F_2}{\partial P_i}\dot P_i  \implies \\
    & K-H = \frac{\partial F_2}{\partial t}, p_i = \frac{\partial F_2}{\partial q_i}, Q_i = \frac{\partial F_2}{\partial P_i}.
\end{split}
\end{equation}

 Hamilton's principle function $S(q,P,t)$ and Hamilton's characteristic function $W(q,P)$ \cite{Mechanics} are both $F_2$ type generating functions. In order to reduce the Hamiltonian to trivial equations of motion, we may consider both of these functions. When the Hamiltonian does not explicitly depend on time, we have that $S(q,P,t) = W(q,p) - T(t)$. In general, the goal is to find the solution to the Hamilton-Jacobi equation such that
\begin{equation}
    H(q_i, \frac{\partial S}{\partial q_i}, t) + \frac{\partial S}{\partial t} = 0.
\end{equation}
This equation only contains partial derivatives and $n+1$ integrable solutions, $\alpha_i$, where $n$ is the degrees of freedom of the system. The procedure is to obtain
\begin{equation}
    S = S(q_1,...,q_n;\alpha_1,...,\alpha_n;t) + \alpha_{n+1}.
\end{equation}
The term $\alpha_{n+1}$ is not of significance. We set $P_i = \alpha_i$, $Q_i = \frac{\partial S}{\partial P_i}$, and solve for $q_i = q_i(Q_k,P_k,t)$, $p_i =  p_i(Q_k,P_k,t)$. In the case that these equations can be solved explicitly, the system is called integrable. To show that the two-body problem is an integrable system, consider the Lagrangian of the system,
\begin{equation}
   L = \frac{m}{2}\left(\dot r^2 + r^2 \dot \varphi^2 \right) - V(r).
\end{equation}
From Hamiltonian Mechanics it follows that
\begin{equation}
    \begin{split}
    &p_r = \frac{\partial L}{\partial \dot r} = m\dot r,\\
    &p_\varphi = \frac{\partial L}{\partial \dot \varphi} = mr^2 \dot \varphi.
    \end{split}
\end{equation}
And the Hamiltonian is
\begin{equation}
    H = p_r\dot r + p_\varphi \dot \varphi + V(r) = \frac{p_r^2}{2m} + \frac{p_\varphi^2}{2mr^2} + V(r).
\end{equation}
And from the Hamilton-Jacobi equation, we obtain
\begin{equation}
\label{eq:HJKepler}
    \frac{1}{2m}\left(\frac{\partial S}{\partial r}\right)^2 + \frac{1}{2mr^2}\left(\frac{\partial S}{\partial \varphi}\right)^2 + V(r).
\end{equation}
We may then set
\begin{equation}
    S(r,\varphi,\alpha_1,\alpha_2,t) = W_r(r) + W_\varphi(\varphi) - \alpha_1t,\; \; \alpha_1 = E.\\
\end{equation}
By substituting into equation \ref{eq:HJKepler}, we obtain
\begin{equation}
\begin{split}
    &\frac{1}{2m}\left(\frac{d W_r}{d r}\right)^2 + \frac{1}{2mr^2}\left(\frac{d W_\varphi}{d \varphi}\right)^2 + V(r) = E \implies\\
    &\frac{1}{2m}\left(\frac{d W_\varphi}{d \varphi}\right)^2 = -\frac{r^2}{2m}\left(\frac{d W_r}{d r}\right)^2 + r^2(E - V(r)).
\end{split}
\end{equation}
It follows
\begin{equation}
    \frac{d W_\varphi}{d \varphi} = \alpha_2.
\end{equation}
We may then set
\begin{equation}
    V_{eff}(r) = V(r) + \frac{\alpha_2^2}{2mr^2}.
\end{equation}
This gives us the result
\begin{equation}
    \frac{dW_r}{dr} = \pm \sqrt{2m(E-V_{eff}(r))}.
\end{equation}
And we obtain
\begin{equation}
    S = \pm \int_{r_0}^r \sqrt{2m(E-V_{eff}(r'))}\;dr' + \alpha_2 \varphi - \alpha_1t.
\end{equation}
Therefore the system is integrable. With that being noted, the goal is now to establish a transformation to preserve the action and keep as many variables cyclic as possible. Specifically we are looking for a transformation $(q_i,p_i) \implies (\theta_i, I_i)$ such that $\dot \theta_i = \frac{\partial H}{\partial I_i} = \omega_i$ and $\dot I_i = -\frac{\partial H}{\partial \theta_i}$ = 0. Hence $H$ should only depend on $I$. We will define our variable to be the area of the phase space enclosed of an orbit divided by $2\pi$ such that 
\begin{equation}
    \begin{split}
    &I = \frac{1}{2\pi} \oint p dq,\\
    &\theta = \omega\cdot t + \delta. 
    \end{split}
\end{equation}
We may consider a general Hamiltonian with no explicit time dependence
\begin{equation}
    E = \frac{p^2}{2m} + V \implies p = \sqrt{2m}\sqrt{E-V(q)}.
\end{equation}
We know that
\begin{equation}
    p = m \frac{dq}{dt} \implies dt = m\frac{dq}{p}  \implies dt = \sqrt{\frac{m}{2}}\frac{dq}{\sqrt{E-V(q)}}.
\end{equation}
A full orbit is then
\begin{equation}
    T = \frac{2\pi}{\omega} = \oint dt = \sqrt{\frac{m}{2}} \oint \frac{dq}{\sqrt{E-V(q)}}.
\end{equation}
This is simply
\begin{equation}
    \frac{d}{dE} \oint \sqrt{2m}\sqrt{V-E(q)}dq = \frac{d}{dE} \oint pdq = 2\pi \frac{dI}{dE}.
\end{equation}
Hence
\begin{equation}
    T = \frac{2\pi}{\omega} = 2\pi \frac{dI}{dE} \implies \frac{1}{\omega} = \frac{dI}{dE}.
\end{equation}
And we have
\begin{equation}
    \frac{dE}{dI} = \frac{dH}{dI} = \omega.
\end{equation}
It then follows 
\begin{equation}
    \theta = \omega \cdot t + \delta.
\end{equation}
Now we may consider the Hamiltonian for the Kepler problem with $V(r) = -k/r$,
\begin{equation}
    H = \frac{p_r^2}{2m} + \frac{p_\varphi^2}{2mr^2} - \frac{k}{r}.
\end{equation}
For the angular motion, the integral is trivial as the angular momentum is constant,
\begin{equation}
    I_\varphi = \frac{1}{2\pi} \int_0^{2\pi} p_\varphi d\varphi = p_\varphi.
\end{equation}
For the radial motion, it is possible to solve for
\begin{equation}
    p_r = \sqrt{2mH - \frac{p_\varphi^2}{r^2} + \frac{2mk}{r}} = \sqrt{2m\left(E+\frac{k}{r}\right)-\frac{p_\varphi^2}{r^2}}.
\end{equation}
It then follows that the resulting integral equation for the radial action variable is 
\begin{equation}
    I_r = \frac{1}{2\pi}\oint p_r dr = \frac{1}{2\pi}2\int_{r_{min}}^{r_{max}}\sqrt{2m\left(E+\frac{k}{r}\right)-\frac{p_\varphi^2}{r^2}}dr.
\end{equation}
The factor of 2 in the integral denotes that in a complete cycle of motion along the plane $r$ cycles from $r_{min}$ to $r_{max}$ and back to $r_{min}$ since the origin is relative to the focal point of an ellipse rather than its center, effectively doubling the path.

We may then note that $k= \lvert E \rvert (r_{max} + r_{min}) $ and and $I_\phi^2 = 2m\lvert E \rvert r_{min}r_{max}$. We then obtain the following
\begin{equation}
    p_r = \sqrt{2m\lvert E \rvert} \frac{\sqrt{(r-r_{min})(r_{max}-r)}}{r}.
\end{equation}
We may then see that
\begin{equation}
    I_r = \frac{\sqrt{2m\lvert E \rvert}}{\pi}\int_{r_{min}}^{r_{max}}\frac{\sqrt{(r-r_{min})(r_{max}-r)}}{r}dr.
\end{equation}
And to solve for the corresponding action variable, we need the following result from \cite{Mechanics},
\begin{equation}
    I = \int_a^b\frac{\sqrt{(r-a)(b-r)}}{r}dr = \pi \left(\frac{a+b}{2} - \sqrt{ab} \right).
\end{equation}
This works out to be the difference between the arithmetic and geometric mean of minimum and maximum distance from the origin. The typical method of solving this integral relies on contour integration, but we can instead consider an argument from \cite{Zurab}, to reduce this integral to an elementary one. First, take $\alpha = (a+b)/2$ and $\beta = \sqrt{ab}$ and parameterize the integral in terms of these new independent variables
\begin{equation}
    I(\alpha,\beta) = \int_a^b\frac{\sqrt{(r-a)(b-r)}}{r}dr = \int_a^b\frac{\sqrt{2r\alpha-r^2-\beta^2}}{r}dr.
\end{equation}
We then obtain
\begin{equation}
    \frac{\partial I}{\partial \alpha} = \int_a^b \frac{dr}{\sqrt{2r\alpha-r^2-\beta^2}} = \int_a^b \frac{dr}{\sqrt{(r-a)(b-r)}}.
\end{equation}
The partial derivative is constant and has no explicit dependence on $\alpha$. Similarly, we may show that
\begin{equation}
     \frac{\partial I}{\partial \beta} = -\int_a^b \frac{\beta }{r\sqrt{2r\alpha-r^2-\beta^2}}dr = -\int_a^b \frac{\sqrt{ab}}{\sqrt{(r-a)(b-r)}}dr.
\end{equation}
Making the substitution $r = ab/s$ then gives
\begin{equation}
    \frac{\partial I}{\partial \beta} = -\int_a^b \frac{\sqrt{ab} }{\sqrt{(ab-as)(bs-ab)}}ds = -\int_a^b \frac{ds}{\sqrt{(s-a)(b-s)}} = -\frac{\partial I}{\partial \alpha}.
\end{equation}
This shows that the partial derivative is constant and has no dependence on $\beta$. It then follows 
\begin{equation}
    I(\alpha,\beta) = \frac{\partial I}{\partial \alpha}\cdot \alpha + \frac{\partial I}{\partial \beta}\cdot \beta + C = \frac{\partial I}{\partial \alpha}(\alpha - \beta) + C.
\end{equation}
for this to be possible, $I$ must be linear in both variables. Now if we assume $\alpha = \beta = a \; \implies I(a,a) = 0 \implies C = 0$. The equation reduces to
\begin{equation}
    I(\alpha, \beta) = \frac{\partial I}{\partial \alpha}(\alpha - \beta).
\end{equation}
To solve for $\frac{\partial I}{\partial \alpha}$, we make the substitution $x = (r-a)/(b-a)$ which implies
\begin{equation}
    \frac{\partial I}{\partial \alpha} = \int_0^1\frac{dx}{\sqrt{x(1-x)}}.
\end{equation}
Using the substitution $x =sin^2{\phi}$ gives us that
\begin{equation}
    \frac{\partial I}{\partial \alpha} = 2\int_0^{\frac{\pi}{2}}d\phi = \pi.
\end{equation}
Therefore
\begin{equation}
    I = \pi\left(\frac{a+b}{2} - \sqrt{ab}\right).
\end{equation}
The solution to the integral equation for the radial variable is then
\begin{equation}
\begin{split}
    I_r &= \frac{\sqrt{2m\lvert E \rvert}}{\pi}\int_{r_{min}}^{r_{max}}\frac{\sqrt{(r-r_{min})(r_{max}-r)}}{r}dr, \\
    &= \frac{\sqrt{2m\lvert E \rvert}}{\pi}\cdot \pi\left(\frac{r_{min}+r_{max}}{2} - \sqrt{r_{min}r_{max}}\right).
\end{split}
\end{equation}

We may then make the substitution that $k/(2\lvert E \rvert) = (r_{max} +r_{min})/2$ and $\sqrt{(I_\varphi^2)/(2m\lvert E \rvert)} = \sqrt{r_{min}r_{max}}$ to obtain
\begin{equation}
    I_r = \sqrt{\frac{m}{2 \lvert E \rvert}}k-I_\varphi \implies E = - \frac{mk^2}{2(I_r+I_\varphi)^2} = H.
\end{equation}
From Hamilton's Equations it follows that
\begin{equation}
\begin{split}
    &\dot \theta_\varphi = \frac{\partial H}{\partial I_\varphi} = \frac{mk^2}{(I_r + I_\varphi)^3},\\
    &\dot \theta_r = \frac{\partial H}{\partial I_r} = \frac{mk^2}{(I_r + I_\varphi)^3}.
\end{split}
\end{equation}
Since the derivatives are constant in time, we obtain
\begin{equation}
\begin{split}
    &\theta_\varphi = \frac{mk^2}{(I_r + I_\varphi)^3}\cdot t + \varphi_0,\\
    &\theta_r = \frac{mk^2}{(I_r + I_\varphi)^3}\cdot t + r_0.
\end{split}
\end{equation}
This shows that the fundamental frequency for $\theta_r$ and $\theta_\varphi$ are identical which implies that the motion is degenerate. This is unique to the $1/r$ potential, and we have previously shown that this broke down under the addition of deterministic perturbation.

\newpage
\chapter{Stochastic Perturbations}
\section{Stochastic Differential Equations}
\label{sec:SDE}
Instead of completely determistic equations, we may consider differential equations subject to stochasticity. In our case we focus on stochastic differential equations resultant from Brownian motion.
To construct Brownian motion, we may consider the process of random walks. First consider some process $R$ with $R_0 = 0$ that is defined as follows
\begin{equation}
    R_n(n) = x_1+\dots+x_n = \sum_{i=1}^n x_i. 
\end{equation}
The $x_i$ are random variables that simply take on the values of $\pm 1$ with $P(x_i = 1) = P(x_i = -1) = 1/2$. This corresponds to a simple symmetric random walk where a particle can jump up or down at discrete time interval, and its position at $t=n$ is given by $R_n$. Each $x_i$ is an i.i.d. Bernoulli random variable with
\begin{equation}
\begin{split}
    &\mathbb{E}(x) = \frac{1}{2}(1 - 1) = 0,\\
    &\Var(x) = \mathbb{E}(x^2) = \frac{1}{2}(1+1) = 1.
\end{split}
\end{equation}
Using the linearity of expectation and variance, we obtain
\begin{equation}
\begin{split}
    &\mathbb{E}(R_n) = 0,\\
    &\Var(R_n) = n.
\end{split}
\end{equation}
We may now define a related random variable $W$ such that $W(0) = 0$ and
\begin{equation}
    W_n(n) = w_1+\dots+w_n = \frac{1}{\sqrt{n}}\sum_{i=1}^n x_i. 
\end{equation}
It then follows
\begin{equation}
    \begin{split}
        &\mathbb{E}(w) = \frac{1}{2}(\frac{1}{\sqrt{n}} - \frac{1}{\sqrt{n}}) = 0,\\
        &\Var(w) = \mathbb{E}(w^2) = \frac{1}{2}(\frac{1}{n}+\frac{1}{n}) = \frac{1}{n}.
    \end{split}
\end{equation}
And similarly, 
\begin{equation}
    \begin{split}
        &\mathbb{E}(W_n(n)) = 0,\\
        &\Var(W_n(n)) = 1.
    \end{split}
\end{equation}
Through the Central Limit Theorem, we then expect that $W_n \rightarrow Z$ with $Z \sim N(0,1)$ in the limit for $n \rightarrow \infty$. This limiting process is called Brownian motion, and if we take our step size $dt = \frac{1}{n}$, we can see that $W_n((k+1)dt) = W_n(kdt) + dW$, where $dW$ is the Brownian increment with:
\begin{equation}
    dW = \pm\sqrt{dt}.
\end{equation}
From the properties of $W$, we can see that
\begin{itemize}
    \item $W_0 = 0$,
    \item $W_t \sim N(0,t)$,
    \item $W_t - W_s = W_{t-s} \sim N(0,t-s)$.
\end{itemize}
Since each $x_i$ is independent each non-overlapping Brownian increment is independent as well. For more information on the properties of Brownian motion see \cite{BM}.

A stochastic differential equation has the form

\begin{equation}
    dX = \alpha(X,t)dt + \beta(X,t)dW.
\end{equation}

There are two main types of stochastic differential equations, which are called Ito and Stratonovich SDE's respectively. While a more thorough overview is given by Gardiner in \cite{Gardiner} we just need to understand that the two forms are equivalent, and that we may convert between an Ito and Stratonovich SDE as outlined in \cite{Gardiner}.

\begin{figure}[ht]%
    \centering
    \subfloat[\centering Euler-Maruyama Method]{{\includegraphics[height = 5cm, trim={0 0 0 0 }, clip]{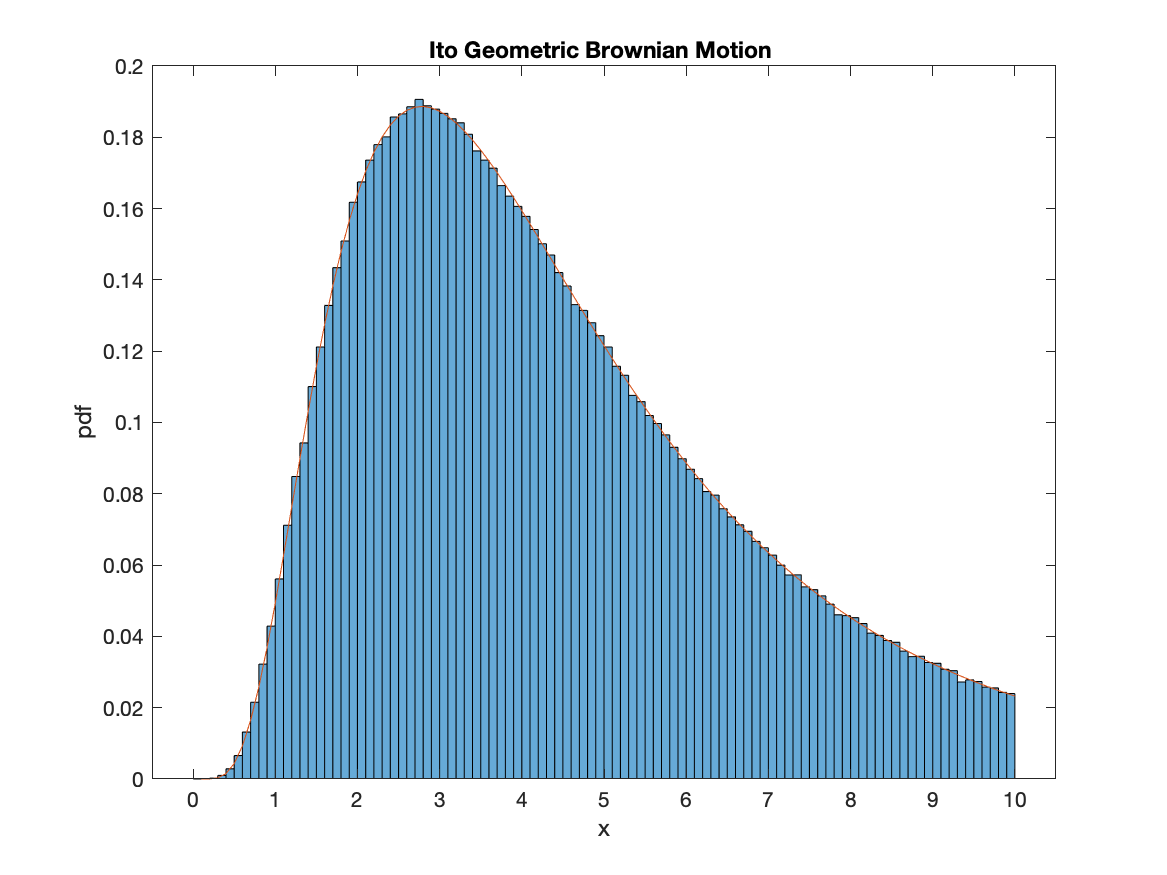}}}%
    \qquad
    \subfloat[\centering Euler-Heun Method]{{\includegraphics[height = 5cm, trim={0 0 0 0 }, clip]{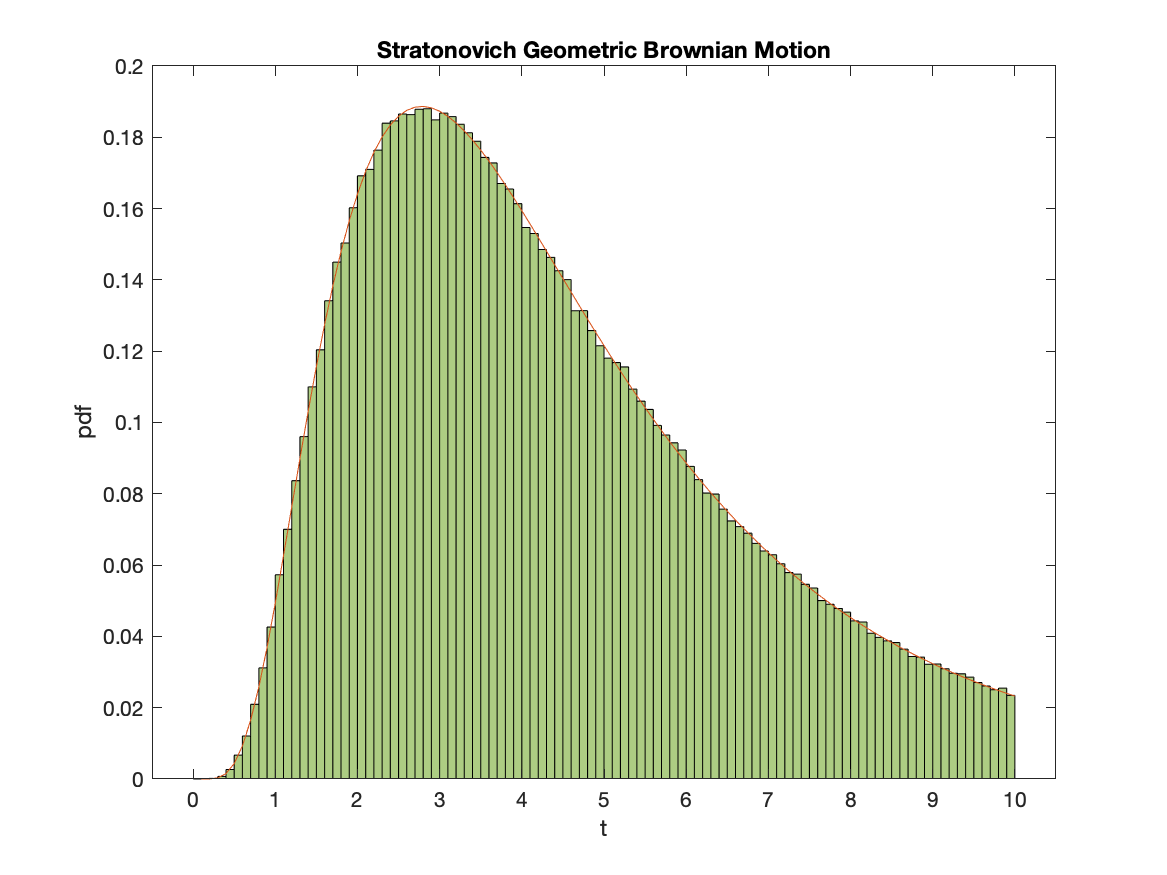}}}%
    \caption{Simulations of geometric Brownian Motion.}%
    \label{fig:GBM}%
\end{figure}

We mainly rely on numerical methods to simulate the resultant motion. There are two main tools that we will consider for this task. There is the Euler-Maruyama method which simulates Ito equation, and there is the Euler-Heun method which simulates Stratonovich equations. A more thorough introduction to these methods and their convergence is given in \cite{SDEInt}. One of the simplest equations with disparate representations we can consider is geometric Brownian motion which has an Ito representation of
\begin{equation}
    dX = \mu X dt + \sigma X dW,
\end{equation}
and a corresponding Stratonovich representation of
\begin{equation}
    dX = \left( \mu - \frac{1}{2}\sigma^2 \right)Xdt + \sigma X dW.
\end{equation}
Geometric Brownian motion has a spatial distribution of
\begin{equation}
    f(t,x) = \frac{1}{\sigma x \sqrt{2\pi t}}\exp{\left(-\frac{(\log x - \log x_0 - \mu t  + \frac{1}{2}\sigma^2t)^2}{2\sigma^2t}\right)}.
\end{equation}
Where $x_0$ is the initial position, $x$ is the final position, and $t$ is the final time. For more information see \cite{GBM}. We can simulate the Ito equation using the Euler-Maruyama method, and we can simulate the Stratonovich equation using the Euler-Heun method. We may then compare the resultant distributions of the final positions with given the probability density function to see how well the methods are performing. The results are outlined in figure \ref{fig:GBM} which were obtained by running 1,000,000 independent geometric Brownian paths using the parameters: $x_0 = 5$, $t = 10$, $\mu = 0$, $\sigma = 1/16$; and binning the final positions using histograms with appropriate scalings to obtain a pdf. 

\begin{figure}[!ht]
    \centering
    \includegraphics[height = 5cm, trim={0 0 0 0 }, clip]{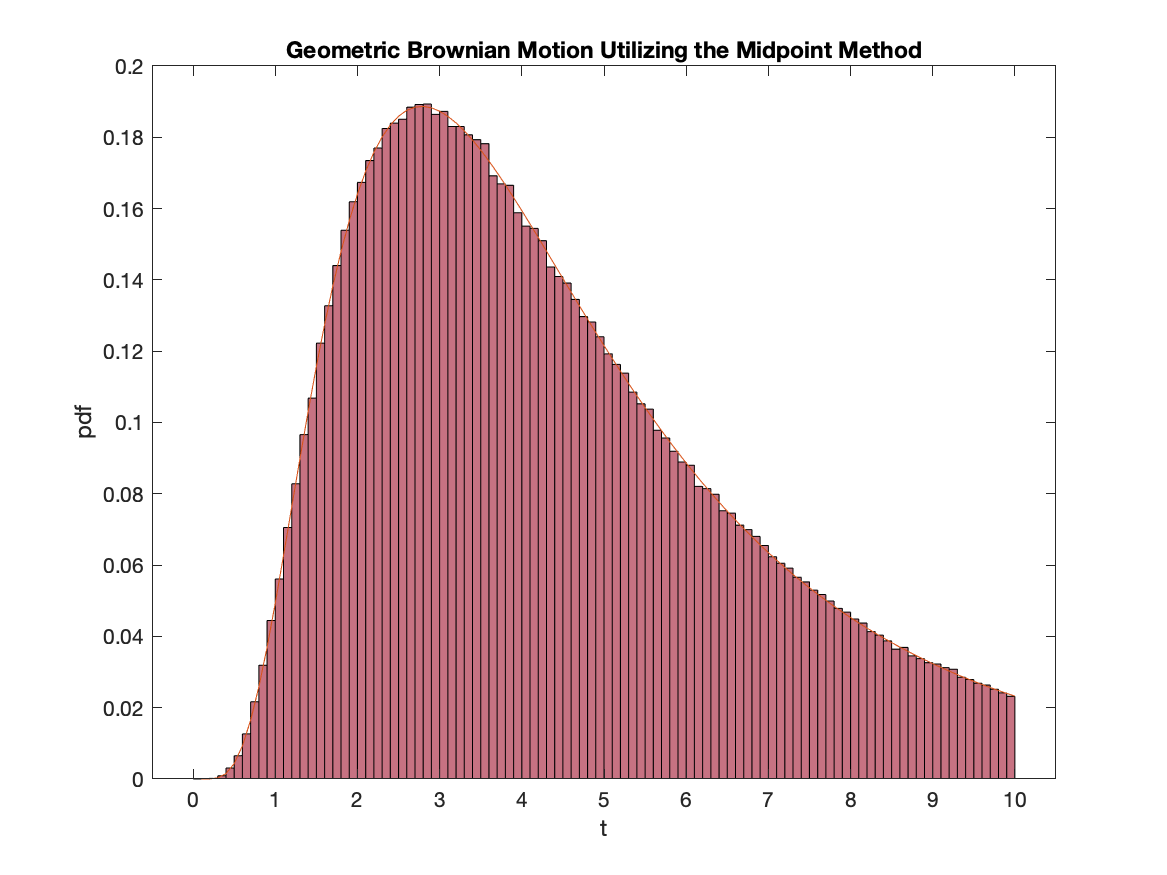}
    \caption{The Midpoint method applied to geometric Brownian motion.}
    \label{fig:midpointGBM}
\end{figure}

\begin{figure}[!ht]
    \centering
    \includegraphics[height = 5cm, trim={0 0 0 0 }, clip]{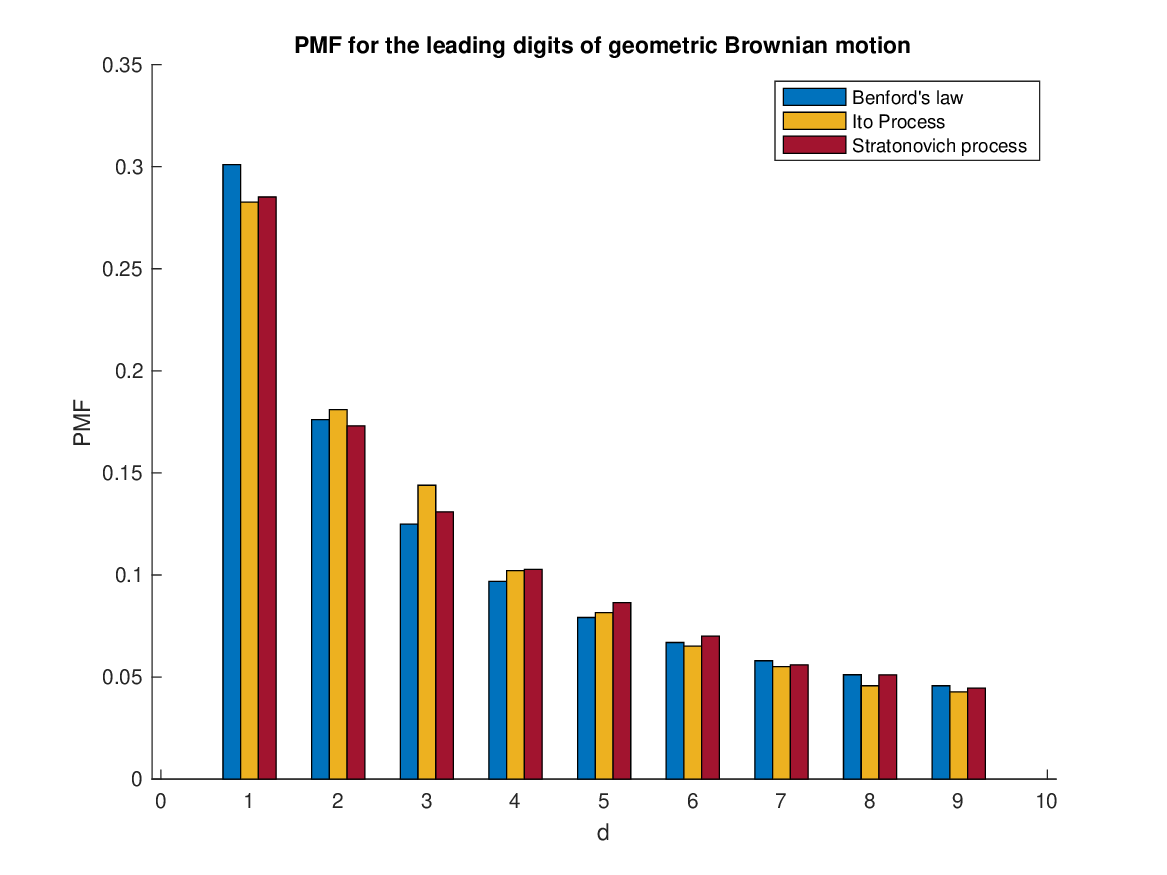}
    \caption{PMF for the leading digits of continuous geometric Brownian motion.}
    \label{fig:BenfordGBM}
\end{figure}

A result of novel interest would be to see how the midpoint method approximates stochastic differential equations. Heuristically this corresponds to using an Ito step to approximate a Stratonovich step, and the method should converge to a Stratonovich process as a result. In a deterministic setting Heun's method and the midpoint method are both RK2 schemes and have equivalent convergence, so we expect similar behavior in the stochastic setting as well. Our results are outlined in Figure \ref{fig:midpointGBM}.
This shows that the two methods behave similarly when simulating geometric Brownian motion. We have not found any literature that analyzes the midpoint method in relation to stochastic differential equations, and a derivation of its convergence and error is outside the scope of this paper, so we leave it as an open problem for future research. 
In a discrete setting, geometric Brownian Motion is a stochastic process whose logarithm is a random walk. A profound result of discrete geometric Brownian motion is that its iterates obey a Benford distribution \cite{MillerTB}. Thorough introductions to Benford's law are provided in \cite{Benford} and \cite{Miller}. For our purposes, we just need to understand that Benford's law defines a logarithmic relationship between the leading digits in certain data sets. In particular, the probability of observing a leading digit, $d$, for a given base, $b$, is
\begin{equation}
    \log_b{\left(\frac{d+1}{d}\right)}.
\end{equation}
Stock prices are typically modeled using continuous geometric Brownian motion, and there have been some results showing that stock returns over obey a Benford distribution over a sufficiently long period of time \cite{StockTurnover}, but there is overall a lack of research into the continous case. That being noted, we can set up simulations by using our numerical methods, and the results are outlined in Figure \ref{fig:BenfordGBM}

It seems plausible that the iterates obey a Benford distribution in the continuous case. A thorough statistical analysis for the data or an analytical proof would have to be the subject of its own paper, but these results may serve as motivation for future research.

\section{Exit Times}
\label{sec:Exit}
We now look at exit times for stochastic processes. Exit times are defined as the minimum time, $\tau$, it takes for the realization of a stochastic process to hit a threshold value. Due to the continuity of these stochastic processes, every point will be hit with certainty, but to simulate a given stochastic process we have to discretize our path, so we must consider the time when our process exceeds a given boundary instead. 

\begin{figure}[ht]%
    \centering
    \subfloat[\centering $h = 10^{-2}$]{{\includegraphics[height = 5cm, trim={0 0 0 0 }, clip]{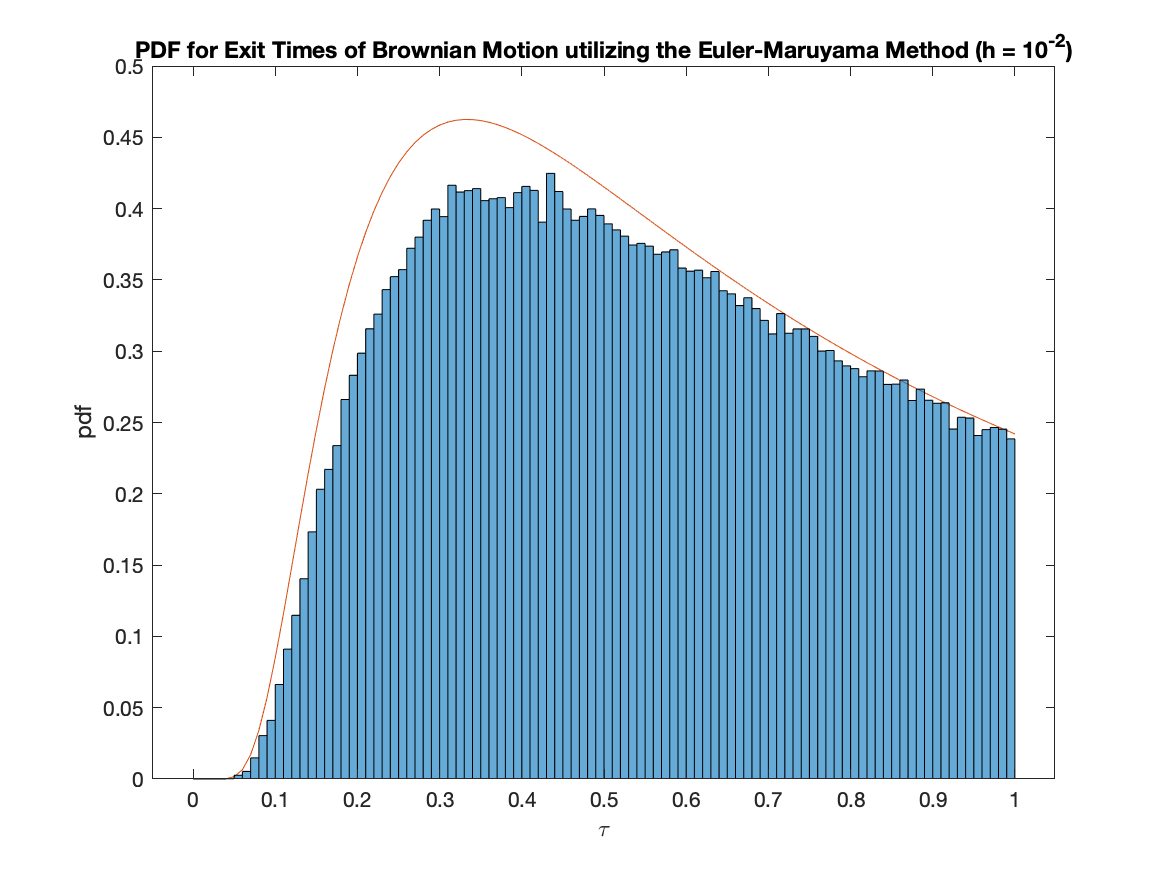}}}%
    \qquad
    \subfloat[\centering $h = 10^{-4}$]{{\includegraphics[height = 5cm, trim={0 0 0 0 }, clip]{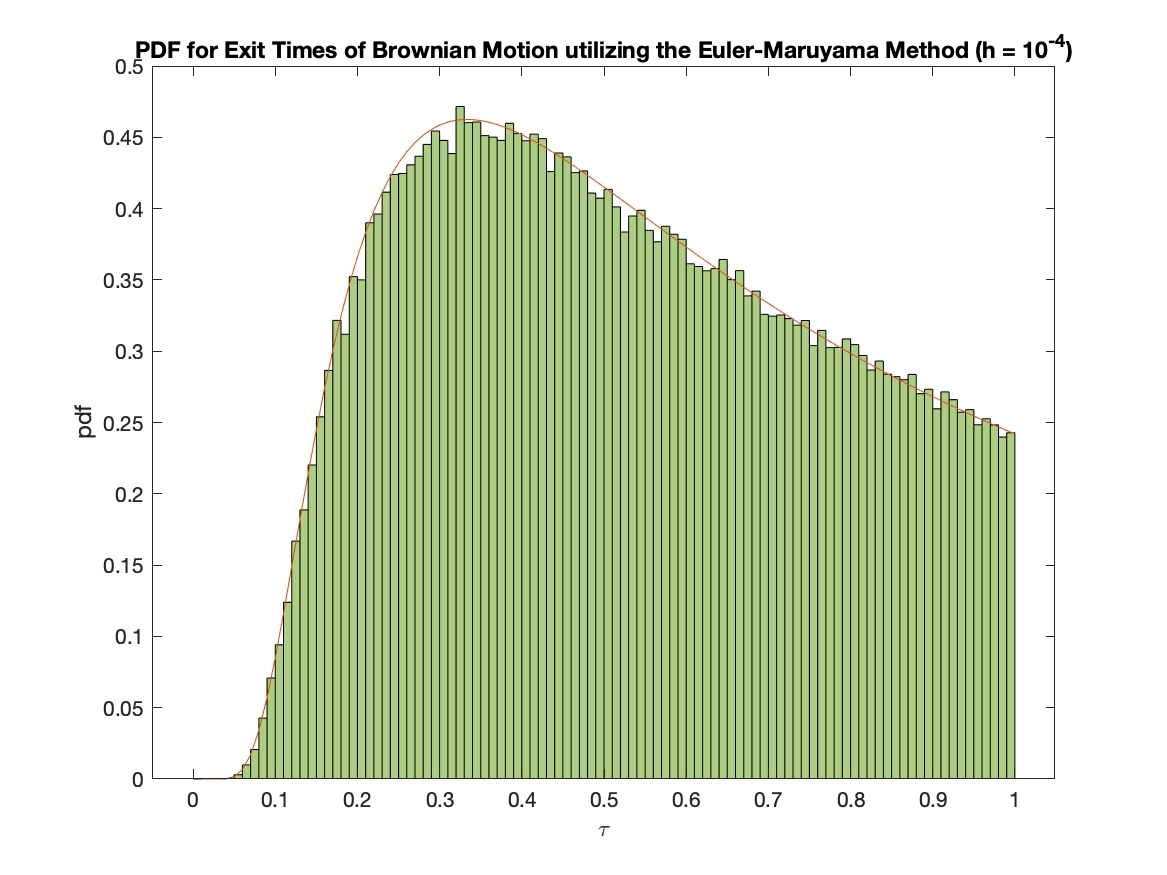}}}%
    \caption{Exit times for Brownian motion utilizing the Euler-Maruyama Method.}%
    \label{fig:ItoBMExitTimes}%
\end{figure}

\begin{figure}[ht]%
    \centering
    \subfloat[\centering $h = 10^{-2}$]{{\includegraphics[height = 5cm, trim={0 0 0 0 }, clip]{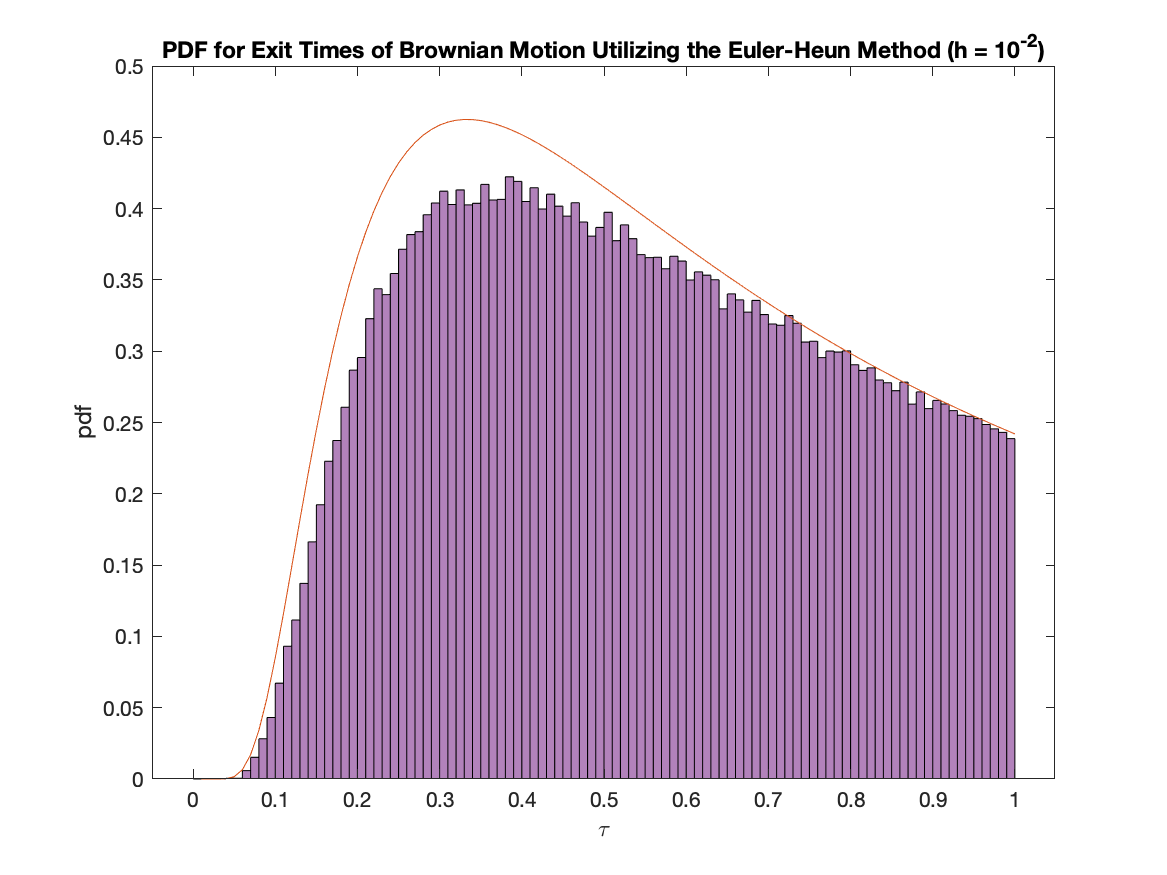}}}%
    \qquad
    \subfloat[\centering $h = 10^{-4}$]{{\includegraphics[height = 5cm, trim={0 0 0 0 }, clip]{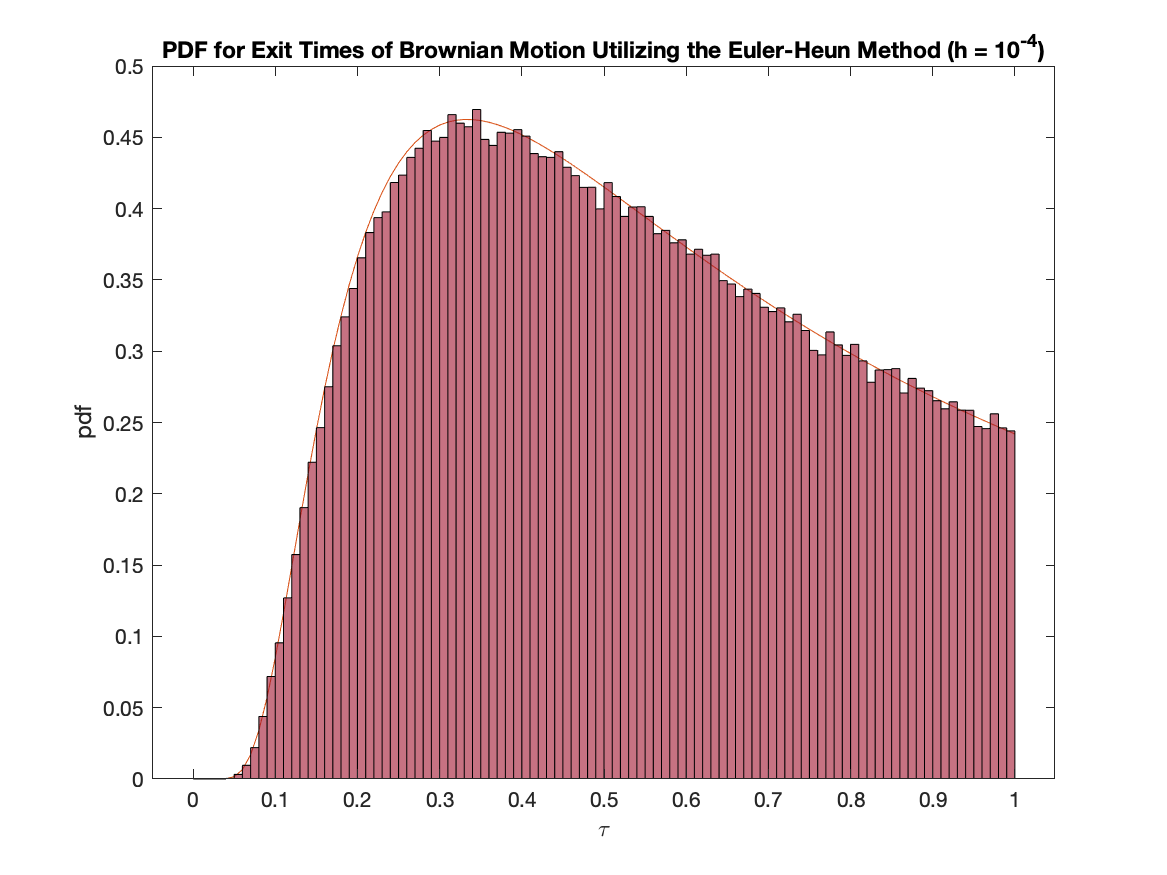}}}%
    \caption{Exit times for Brownian motion utilizing the Euler-Heun Method.}%
    \label{fig:StratBMExitTimes}%
\end{figure}

Exit times play an important role in many stochastic processes, and the simplest case is to consider the exit time of Brownian motion. The cdf for the exit time of the interval $(-\infty, x)$ for Brownian motion is given by

\begin{equation}
    P(\tau \geq t) = 2\left(1 - \Phi\left(\frac{x}{\sqrt{t}}\right)\right). 
\end{equation}

where $\Phi$ is the cdf for the standard normal distribution \cite{BM}. When simulating Brownian motion using either of our numerical methods method we obtain the following results outlined in Figures \ref{fig:ItoBMExitTimes} and \ref{fig:StratBMExitTimes}.

In the deterministic setting, the Euler-Maruyama method and Euler-Heun method correspond to RK1 and RK2 schemes respectively, which implies that they have first and second order convergence respectibely. In this stochastic setting both methods have strong convergence with order 1 and weak convergence with order 1/2 (see \cite{SDEInt}). In the case of purely additive noise, both methods are degenerate cases of the Milstein method with converges both strongly and weakly with order 1 (see \cite{SDEInt}). Even with the higher convergence due to the purely additive noise, extremely small step sizes are needed for convergence since results for exit times depend on the continuity of the sample paths and time intervals, but simulating the motion naturally requires discretization.

The numerics are still sufficient for our purposes, as there are very few practical higher order methods for stochastic equations. There has been interest in using neural networks to simulate stochastic differential equations \cite{NNSDE}, and it would be worthwhile to study the convergence of these machine learning models. It may also be possible to leverage results from the field of topological data analysis \cite{TDA} to study the topology of these neural networks or devise new methods to analyze stochastic processes. This would require a dedicated study, so for our purposes we will stick to the Euler-Heun and Euler-Maruyama method, but we must be mindful of their convergence issues. It is also worth noting that since the Euler-Maruyama and Euler-Heun methods have identical convergence, we may use whichever is more suitable for the relevant stochastic process.

The resultant distribution for the exit times of Brownian motion has an infinite expectation. Due to its independent and continuous increments, a Brownian path will hit every point with certainty, but we cannot predict how long it will take on average to hit any given point. More complex stochastic processes can have finite exit times, and one of the most well known examples is simply a Bessel process. We have the process
\begin{equation}
    X_t = \sqrt{W_{1t}^2 + W_{2t}^2}.
\end{equation}
The $W_{it}$ are independent Brownian motions. By setting $X_t = f(W_{1t},W_{2t})$, we can use the two-dimensional version of Ito's Lemma outlined in \cite{Gardiner} to derive the corresponding stochastic differential equation for this process, and as a result we obtain
\begin{equation}
    dX_t = \frac{\partial f}{\partial W_{1t}}dW_{1t} + \frac{\partial f}{\partial W_{2t}}dW_{2t} +
    \frac{1}{2}\left(\frac{\partial^2 f}{\partial W_{1t}^2} + \frac{\partial^2 f}{\partial W_{2t}^2}\right)dt.
\end{equation} 
This simplifies to
\begin{equation}
    dX_t = \frac{1}{2X_t}dt + dW_t.
\end{equation}
Where 
\begin{equation}
    dW_t = \frac{W_{1t}}{\sqrt{W_{1t}^2 + W_{2t}^2}}dW_{1t} + \frac{W_{2t}}{\sqrt{W_{2t}^2 + W_{2t}^2}}dW_{2t}.
\end{equation}
The term $dW_t$ is simply Brownian motion as the combination of two independent Brownian paths is simply a new Brownian path. We may now imagine starting the process in some interval $(a,b)$ at some point $x$ with absorbing boundary conditions. We to find the probability density function for the spatial evolution of our system, in particular we wish to find $p(x',t|x,0)$ which is the pdf that the system evolves to $(x',t)$ from $(x,0)$. This pdf satisfies the Kolmogorov backwards equation \cite{Gardiner}. Ifour Bessel process starts at $r$ and exits at $R_1$ with $r < R_1$ with the condition that $p(R_1) = 0$, the mean first passage time $T$ satisfies the partial differential equation
\begin{equation}
    -1 = \frac{1}{2r}T_{r} + \frac{1}{2}T_{rr}.
\end{equation}
This has the solution
\begin{equation}
    T = \frac{1}{2}(R_1^2-r^2). 
\end{equation}
Therefore, starting from an initial height of $R_0$ the mean time required to hit a radius $R_1$ for a Bessel process is given by
\begin{equation}
    \mathbb{E}(\tau) = \frac{R_1^2-R_0^2}{2}.
\end{equation}

Utilizing the Euler-Heun method with $R_0 = 0$ and $R_1 = 1$, we obtained the following distribution from 1,000,000 stochastic paths

\begin{figure}[ht]
    \centering
    \includegraphics[height = 5cm, trim={0 0 0 0 }, clip]{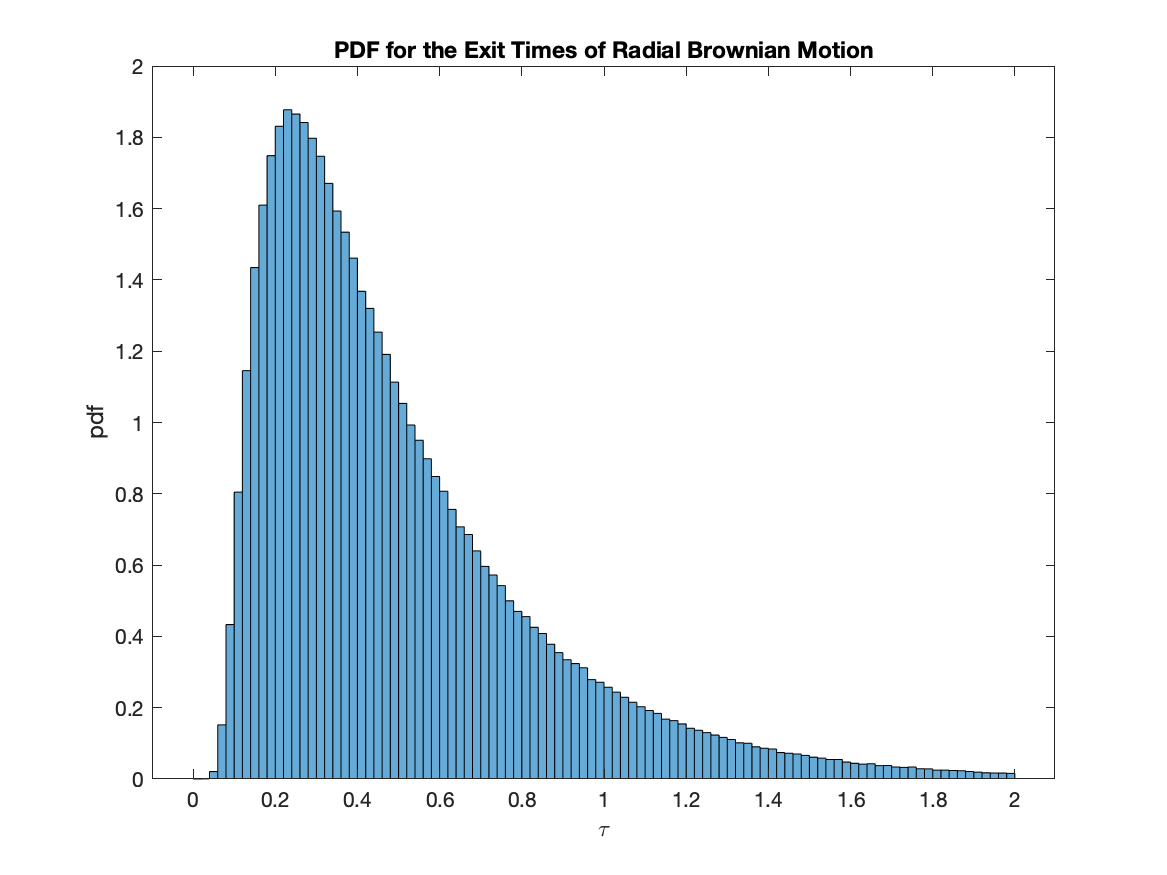}
    \caption{PDF for the exit times of radial Brownian Motion.}
    \label{fig:RadialBMExitTimes}
\end{figure}

From this histogram we can compute an expected value of 0.4841 which is sufficiently close to the analytical value of 0.5 to give us confidence in our numerical methods for more complex stochastic processes. 


\section{The Stochastic Kepler Problem}
\label{sec:SKP}
Going forward we are interested in verifying the work outlined in  "Soliton Perturbations and the Random Kepler Problem" \cite{KeplerExit}. Under certain conditions,  the propagation of solitons in an optical fiber are fundamentally similar to the orbits of the stochastically perturbed Kepler problem. The mean exit time for these stochastically perturbed Kepler potentials then corresponds to the mean distance before dissipation of a soliton in an optical fiber. Though the technical details are beyond the scope of this paper, going forward we are interested in verifying the methodology of the paper and ideally obtaining numerical results to compare with the analytical solutions presented. 

To begin, let us now consider a Hamiltonian system with action-angle variables, $(I,\phi))$ and an initial Hamiltonian of $H_0(I)$. We may consider random perturbations that preserve the Hamiltonian structure which gives us a perturbed Hamiltonian as follows
\begin{equation}
    H = H_0 + \epsilon\gamma(t)V(\phi,I).
\end{equation}
Where $\epsilon$ is a vanishingly small timescale factor, $\gamma(t)$ is standard Gaussian white noise which can be thought of as the realization of Brownian motion with some time correlation, and $V(\phi,I)$ is a stochastic potential function. The use of Gaussian white noise preserves the Hamiltonian structure, so the resulting equations of motion are 
\begin{equation}
    \begin{split}
        &\dot I = -\frac{\partial H}{\partial \phi} = -\epsilon\gamma(t)\frac{\partial V}{\partial \phi},\\
        &\dot \phi = \frac{\partial H}{\partial I} = \omega + \epsilon\gamma(t)\frac{\partial V}{\partial I}.
    \end{split}
\end{equation}
This can be rewritten as follows,
\begin{equation}
        \begin{split}
        &dI = -\epsilon dW\frac{\partial V}{\partial \phi},\\
        &d\phi =  \omega dt + \epsilon dW\frac{\partial V}{\partial I}.
    \end{split}
\end{equation}
Going forward, we will utilize the methodology outlined in \cite{KeplerExit}. To this end, we assume that our Brownian increments have a time correlation $2\delta(t-s)$ which corresponds to multiplying our typical Brownian increments by a factor of $\sqrt{2}$. We may also assume that this is a Stratonovich process and that the evolution of the probability density function is given by the Stratonovich form of the Fokker-Planck equation with the exact calculations presented in equation 30 of \cite{KeplerExit}. We may then average over the periodic orbits to remove dependence on the angular term. It is worth noting that the results depend on $\epsilon$ being vanishingly small, and that there are naturally computational limitations on how small we can make $\epsilon$ which will naturally introduce additional error.

To verify this methodology, we may consider the example of the time it takes for a sufficiently small action to escape a finite interval $(\alpha, \beta)$ starting from some initial conditions $(\phi_0, I_0)$. For small action, the Kepler problem can be well approximated as the harmonic oscillator. We may then use the action angle representation of the harmonic oscillator outlined in \cite{HarmOsc} which is given by
\begin{equation}
\begin{split}
    &H_0 = I \omega,\\
    &q = \sqrt{\frac{2I}{m\omega}}\cos{\phi},\\
    &p = \sqrt{2mI\omega}\sin{\phi}.
\end{split}
\end{equation}
We may take our stochastic potential as
\begin{equation}
    V = \frac{q^2}{2} = \frac{I}{m\omega_0}\cos^2{\phi}.
\end{equation}
The resultant modulation equations for the modified Kepler problem are
\begin{equation}
    \begin{split}
        &dI = \frac{2\epsilon I}{m\omega_0}\cos{\phi}\sin{\phi}dW,\\
        &d\phi = \omega(I) dt + \frac{\epsilon}{m \omega_0}\cos^2{\phi}dW.
    \end{split}
\end{equation}
Because we are averaging over the orbits of the Kepler problem, and since we are only interested in the radial diffusion, we can use a constant $\omega$ for our simulations of the angular drift term. We can consider more complicated cases with varying parameters, but this should provide a sufficient approximation for a finite interval sufficiently far enough from the separatrix.

\begin{figure}[ht]
    \centering
    \includegraphics[height = 5cm, trim={0 0 0 0 }, clip]{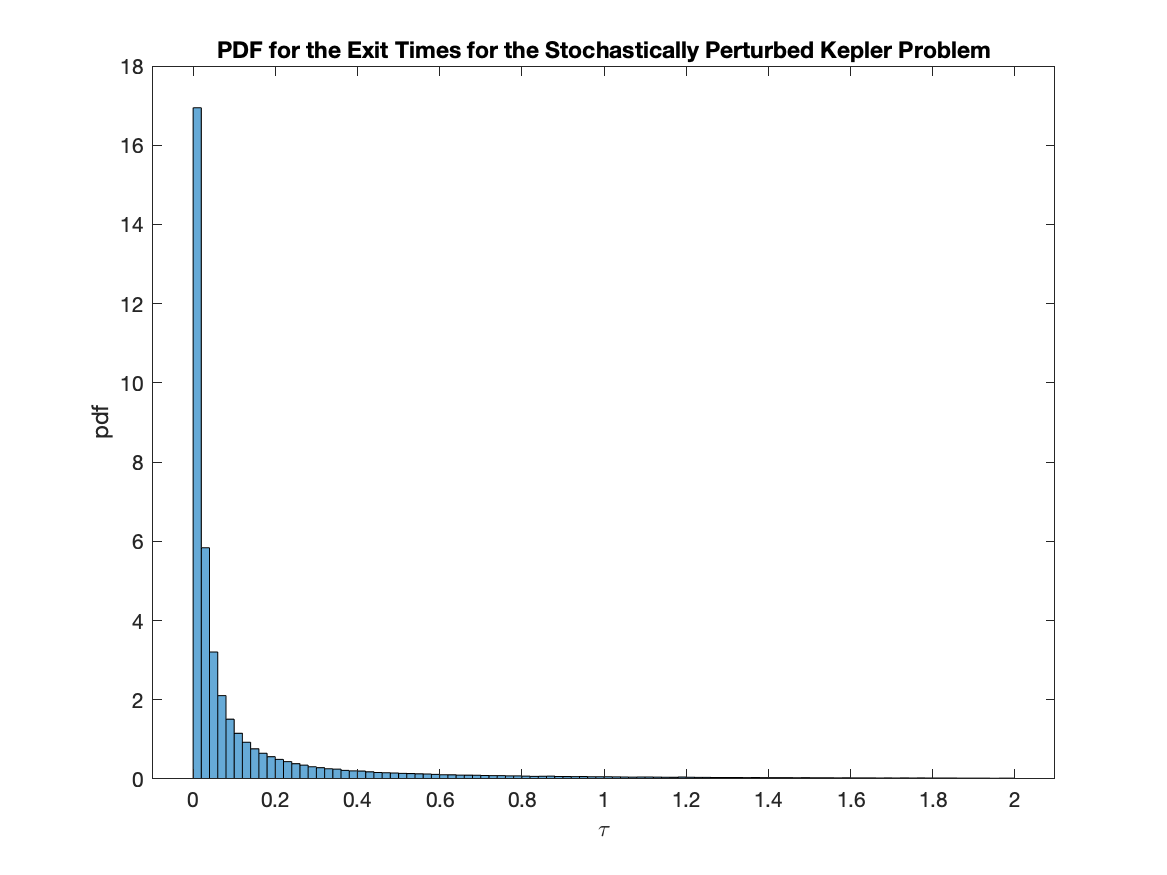}
    \caption{PDF for the exit times of the stochastically perturbed Kepler Problem.}
    \label{fig:KeplerEscapeTime}
\end{figure}

We may then then calculate
\begin{equation}
    \begin{split}
    &A(I) = \frac{1}{2\pi}\int_0^{2\pi}V_{\phi}^2d\phi = \frac{I^2}{2m^2\omega^2},\\
    &\int\frac{d\eta}{A(\eta)} = -2m^2\omega^2\eta^{-1},\\
    &\int\frac{\eta d\eta}{A(I)} = 2m^2\omega^2\log{\lvert \eta \rvert }.
    \end{split}
\end{equation}
By utilizing equation 35 in \cite{KeplerExit} we may obtain the following result for the mean exit time
\begin{equation}
    \mu_{(\alpha,\beta)} = \frac{2m^2\omega^2}{\epsilon^2}\frac{(I^{-1}-\alpha^{-1})(\log{\frac{\beta}{\alpha}})-(\log{\frac{I}{\alpha}})(\beta^{-1}-\alpha^{-1})}{\beta^{-1}-\alpha^{-1}}.
\end{equation}
It is worth noting that the mean time to infinity does not exist for this potential, as the integral for this exit time outlined in equation 39 of \cite{KeplerExit} diverges
\begin{equation}
    \mu_{\infty} = \int_{I_0}^\infty \frac{IdI}{A(I)} = \int_{I_0}^\infty \frac{2m^2\omega^2}{I}dI = \infty.
\end{equation}
We are now interested in verifying our result using our numerical methods. We initialized $m = 1000000$ parallel stochastic paths to run our simulations. We used the parameters $I_0 = 7.5$, $\alpha = 5$, $\beta = 10$, $\omega = 0.01$, $m = 0.01$, and $\epsilon = 0.0001$. As we assumed this is a Stratonovich process, we utilize the Euler-Heun method, and to simulate the effect of averaging out the angular dependence, we may then consider partitioning $[0,2\pi)$ into $m$ distinct intervals, and then using those values as a vector of initial conditions to help remove dependence on our angular term. We may then bin and the exit times into histograms and scale them accordingly. The resulting distribution is outlined in Figure \ref{fig:KeplerEscapeTime}.
The results give a mean exit time of .1240 compared with the analytical result of 0.1133, which shows a strong agreement between the analytical results and the simulations and that the methodology in calculating these exit times is sound. We could also obtain the exact distribution by solving the Fokker-Planck equation, but the methods required are outside the scope of this paper.
\newpage
\chapter{Conclusion}
\section{Results}
We investigated the effects of specific perturbations in the Kepler problem. We observed that the addition of a deterministic potential derived from the first order solution to the Einstein field equations causes the standard elliptical orbits observed under Newtonian gravity to form precessions. We were able to fit the parameters of our model to account for the anomalous precession of Mercury, which was one of the famous tests for General Relativity. We then studied stochastic perturbations by deriving an analytical solution for the mean passage time for a certain stochastic potential in the Kepler problem. We verified our analytical solution by utilizing Monte Carlo simulations, and this also serves as a demonstration of the methodology for more complex potentials outlined in "Soliton Perturbations
and the Random Kepler Problem" \cite{KeplerExit}.

\section{Future Work}
The work in "Soliton Perturbations
and the Random Kepler Problem" \cite{KeplerExit} outlines the mean exit times for the action of the Kepler problem to infinity under certain potentials. The natural progression of our work would be to verify the results they outline in equations 41 and 47. Due to practical and technical limitations, we cannot provide simulated data to check their results. On the hardware we had available, it took 8 hours of computation to calculate the results for the harmonic oscillator approximation. The equations for the exit times to infinity are much more complicated and we do not have a time scale for how long it would take to calculate; it also must be noted that the equations are not well-behaved near the origin and break down near the separatrix of the Kepler problem, and the results also depend on $\epsilon$ being a vanishing term. These factors combined with the convergence of our numerical methods would make interpreting results very difficult. Perhaps with more extensive hardware or more refined methods, simulating the results would be practical. Still, with the methodology verified by our work, anyone with the proper interest and means may now look at verifying those calculations as the problem has stood for over 20 years as of the writing of this paper.

\printbibliography
Jesse Dimino, Department of Mathematics, CUNY College of Staten Island, Staten Island, NY 10314 {\textcolor{blue}{\href{mailto:jessedimino@gmail.com}{jessedimino@gmail.com}},
\textcolor{blue}{\href{mailto:jesse.dimino@macaulay.cuny.edu}{jesse.dimino@macaulay.cuny.edu}}}

\end{document}